\documentclass{article}

\usepackage{arxiv}

\usepackage[utf8]{inputenc} 
\usepackage[T1]{fontenc}    
\usepackage{hyperref}       
\usepackage{url}            
\usepackage{booktabs}       
\usepackage{amsfonts}       
\usepackage{nicefrac}       
\usepackage{microtype}      
\usepackage{lipsum}
\usepackage{graphicx}
\usepackage[numbers]{natbib}
\usepackage{amsmath}
\usepackage{mathtools}
\usepackage{amssymb}
\usepackage{tabularx}
\usepackage[table,xcdraw]{xcolor}
\graphicspath{ {./images/} }

\title{Formation of Comets}

\author{
 Jürgen Blum \\
  Institute for Geophysics and Extraterrestrial Physics\\
  Technische Universität Braunschweig\\
  Mendelssohnstraße 3, 38106 Braunschweig, Germany\\
  \texttt{j.blum@tu-braunschweig.de} \\
   \And
 Dorothea Bischoff \\
  Institute for Geophysics and Extraterrestrial Physics\\
  Technische Universität Braunschweig\\
  Mendelssohnstraße 3, 38106 Braunschweig, Germany\\
  \And
 Bastian Gundlach \\
  Institute for Geophysics and Extraterrestrial Physics\\
  Technische Universität Braunschweig\\
  Mendelssohnstraße 3, 38106 Braunschweig, Germany\\
}

\begin{document}
\maketitle
\begin{abstract}
Questions regarding how primordial or pristine the comets of the solar system are have been an ongoing controversy. In this review, we describe comets' physical evolution from dust and ice grains in the solar nebula to the contemporary small bodies in the outer solar system. This includes the phases of dust agglomeration, the formation of planetesimals, their thermal evolution and the outcomes of collisional processes. We use empirical evidence about comets, in particular from the Rosetta Mission to comet 67P/Churyumov--Gerasimenko, to draw conclusions about the possible thermal and collisional evolution of comets.
\end{abstract}


\section{Introduction}
\label{sec:Introduction}

\begin{figure}

\centering
\includegraphics[width=13.79cm]{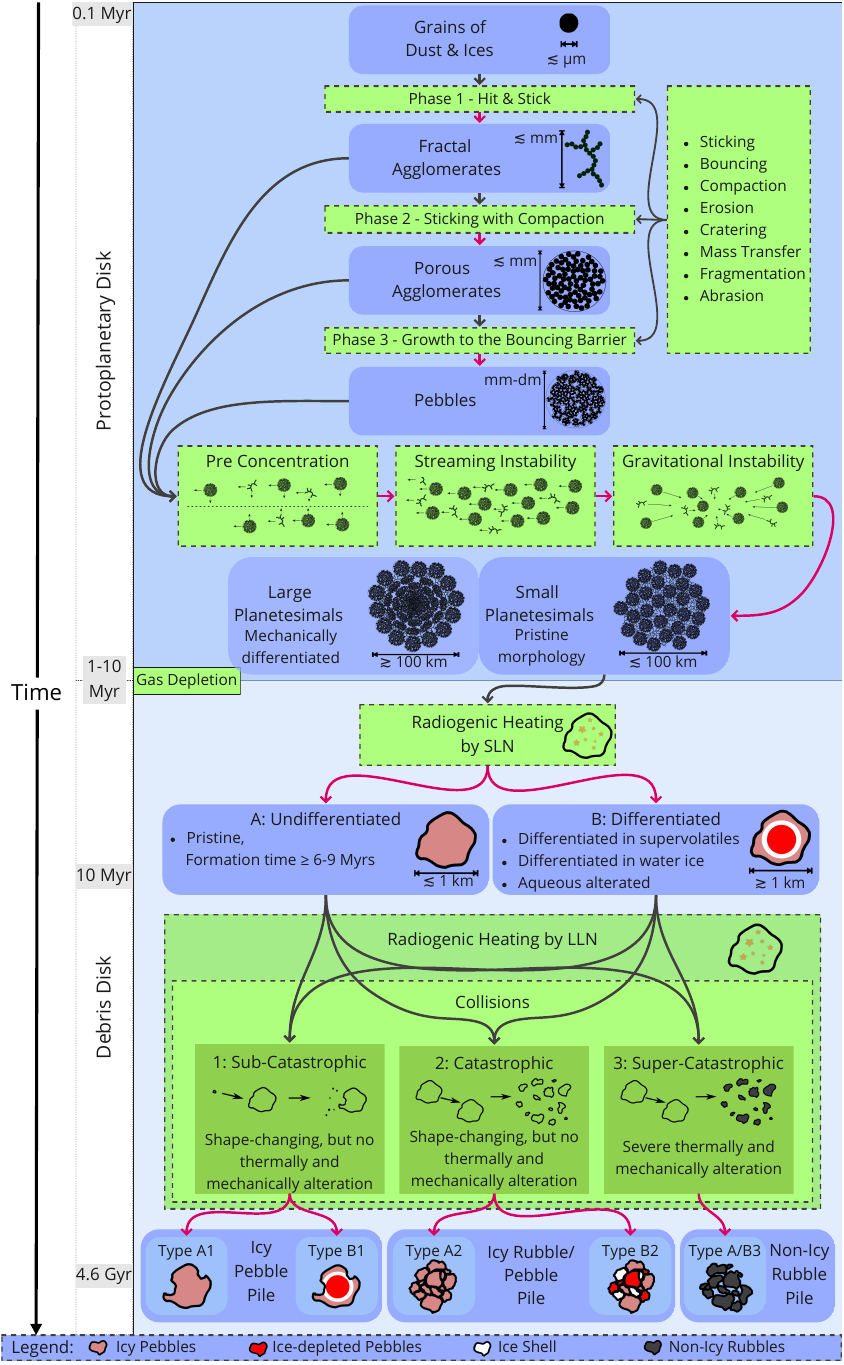}
\caption{This overview is a graphical representation of the formation and evolution of planetesimals beyond the H$_2$O snowline. The~blue and green backgrounds denote objects and processes, respectively. Each step is explained and discussed in the paper. A~detailed version can be found at 
 \url{https://www.tu-braunschweig.de/fileadmin/Redaktionsgruppen/Institute_Fakultaet_5/IGEP/AG_Blum/Comet_Formation_detailed.pdf} (accessed on 5 May 2022) 
 and in the Supplementary~Materials. \label{fig:overview}}
\end{figure}

Comets are believed to be the most pristine objects of our solar system. They consist of several ices, mainly water ice but also super-volatiles such as CO or CO$_2$ ice, as~well as refractory materials such as minerals, organics and salts. This composition obviously results in cometary gas and dust if the insolation is intense enough. In~principle, a~comet becomes dust-active if the outgassing rate is large enough to release dust particles against holding forces such as gravity and cohesion~\cite{Kuhrt.1994}. However, the~exact physical mechanism of the ejection of dust is not yet well understood. Additionally, several aspects of the formation and evolutionary pathways of comets are still under debate. In~this work, we provide a detailed overview of how icy planetesimals may have formed in the protoplanetary disc (PPD) and may have evolved into present-day bodies of the solar system, including comets. In~Figure~\ref{fig:overview}, we provide a graphical overview of the formation and evolutionary stages from protoplanetary dust to the contemporary small bodies in the solar system; we explain the stages in detail in the following sections. A~more detailed version of \mbox{Figure~\ref{fig:overview}} can be found at \url{https://www.tu-braunschweig.de/fileadmin/Redaktionsgruppen/Institute_Fakultaet_5/IGEP/AG_Blum/Comet_Formation.pdf}  
(accessed on 5 May 2022) and in the Supplementary Materials. In~Section~\ref{sec:Coagulation}, the~coagulation processes of the dust and ice particles are summarised, beginning with an overview of protoplanetary discs. In~this work, we use the nomenclature proposed by Güttler et al. 2019 \cite{Guttler.2019}. 
When referring to particles, their sizes are not further constrained. However, (dust or ice) grains are restricted to the (sub-)micrometre size range and are assumed to be homogeneously composed of one material only. Agglomerates consist of grains and can be heterogeneous in composition. The~hit-and-stick growth of particles stops at growth barriers where pebbles, millimetre- to decimetre-sized porous agglomerates, are formed. Section~\ref{sec:Formation_of_Planetesimals} discusses the formation of planetesimals from pebbles via streaming instability and subsequent gravitational collapse. The~required properties of the particles and the protoplanetary disc, as~well as other scenarios, are briefly discussed. The~different evolutionary processes that can alter the formed planetesimals are then presented in Section~\ref{sec:Evolution}. The~formation processes in combination with evolutionary alterations result in three different categories of evolved planetesimals: icy pebble piles, icy rubble/pebble piles and non-icy rubble piles, which can be further subdivided depending on the evolutionary processes. In~Section~\ref{sec:Discussion}, we discuss the properties of possible contemporary end products of the evolution and compare them to observations of comets in the solar system to connect them to the formation and evolution pathways. Finally, we summarise the paper and discuss open questions in Section~\ref{sec:Conclusion}. 

It must be emphasised that we only provide one plausible, but~not the only possible scenario for the formation of planetesimals and evolution into comets. {This model heavily relies on the existence of pebbles, for~which only indirect evidence exists~\cite{Blum.2017}.} \color{black} Alternative {planetesimal-formation} \color{black}models exist and are discussed in detail by Weissman et al. 2020 \cite{Weissman.2020}.

\section{From Dust To Pebbles}
\label{sec:Coagulation}
\subsection{Protoplanetary discs}
\label{sec:PPD}
Directly connected to the formation of a star from the gravitational collapse of a molecular cloud is the formation of a protostellar disc, which then evolves into a PPD within \mbox{$\sim$$10^4$ years}~\cite{Yorke.1993,Hueso.2005}. This process is initially connected to high temperatures, but~while material from the disc gets accreted by the star, temperature and mass decrease. This transition from a protostellar to a protoplanetary disc lasts about $\sim$$0.5$ Myrs for a solar-type star and is determined by the complete dispersion of the envelope, resulting in a disc mass of a few percent of the mass of the central star~\cite{Williams.2011}. The~following evolution of the disc is dominated by the ongoing accretion of material onto the star, photoevaporation, dust agglomeration and dynamical interactions with the stellar or galactic environment. Properties of protoplanetary discs are traditionally observed by optical or near-IR methods. With~the Atacama Large Millimeter/submillimeter Array (ALMA), a~new era of observation of protoplanetary discs has started. ALMA has shown, due to its high angular resolution and sensitivity, detailed substructures of PPDs, namely spirals, gaps, axisymmetric rings and inner cavities~\cite{Andrews.2018,Huang.2018,Long.2018}, which were unexpected from the traditional view of PPDs~\cite{Liu.2020}. Several ideas have been proposed to explain these structures, e.g.,~interaction with planetary bodies, which are beyond the scope of this review. Protoplanetary disc masses can be constrained by ALMA measurements using different methods. For~example, the~mass of solids in millimetre-sized dust grains can be estimated from the millimetre flux density combined with assumptions about the optical depth, dust opacity and disc temperature (e.g., \cite{Barenfeld.2016,Ansdell.2016,Pascucci.2016,Ansdell.2017,Cieza.2019,Mulders.2021}). Another method uses the amplitude and wavelength of gravitational instability wiggles~\cite{Terry.2021}. The~total dust masses and  typical PPD sizes seem to correlate, as~low-mass discs are generally smaller than massive discs~\cite{Tripathi.2017,Andrews.2018,Hendler.2020}.
\par When the temperatures inside the PPD have sufficiently dropped, submicrometre- to micrometre-sized solid particles start to condense. The~condensation sequence depends on the local temperature and pressure in the disc (see \cite{Li.2020b} for a list of condensation temperatures for several elements). Complex 2D models describe the chemical evolution of PPDs and the different material compositions in different regions of the disc (e.g., \cite{Pignatale.2016} and references therein). For~example, forsterite, enstatite, olivine, pyroxene and quartz are important minerals, whose abundances can be compared to meteoritic and cometary dust compositions (e.g., Figure~11 in~\cite{Gail.2004}).  
\par
For water ice, the~critical temperature of condensation is between $145 \,\mathrm{K}$ \cite{Podolak.2004} and \mbox{$170 \,\mathrm{K}$ \cite{Hayashi.1981}}, depending on the assumed local pressure. However, the~locations of the snowlines of various ice species can vary due to several processes in the disc~\cite{Martin.2012,Panic.2017,Xiao.2017}. It has also been argued that the condensation of volatile gases onto refractory grains at the position of the respective snowlines can be a possible growth mechanism for solid bodies~\cite{Ros.2013}. The~processes acting on dust and gas in PPDs are key elements to understand planetesimal formation. While gasses (mostly H$_2$ and He) dominate the mass of the PPD, the~dust is essential for the formation of the first larger solid bodies. The~effects behind dust growth will be presented in the following. 
\par The orbital gas velocity in PPDs is reduced below the Keplerian speed because the negative radial gas-pressure gradient counteracts the gravitational pull of the star. Because~solid material is not pressure-supported, dust particles should orbit the central star with Keplerian velocity. However, as~they experience a headwind due to the slower-rotating gas, a~drag force acts on the particles. This drag can be described by the Epstein drag law for particles that are smaller than the mean free path of the gas. For~particle size exceeding the mean free path of the gas, the~Stokes regime becomes relevant for the gas drag. The~gas drag acting on dust particles is best described by the stopping time, which can be written as
\begin{equation}
\tau_f = \begin{cases}
\frac{R \, \rho_p}{c_s \, \rho_g} & \text{in the Epstein regime}, \\ \frac{4 \, \rho_p \, R^2}{9 \, c_s \, \rho_g \lambda} & \text{in the Stokes regime.}
\end{cases}
\label{eq:tauf}
\end{equation}

Here, $R$, $\rho_p$, $c_s$, $\rho_g$ and $\lambda$ are the particle radius, particle density, gas sound speed, gas density and mean free path, respectively. The~Stokes number as a common measure of particle sizes in PPDs is defined as
\begin{equation}
    St = \Omega \, \tau_f \,
\end{equation}
with the Keplerian frequency $\Omega$. Particles with the same Stokes number behave aerodynamically similar. For~a given PPD model and a fixed heliocentric distance, the~Stokes number corresponds to a specific 
value of $R \, \rho_p$ or $R^2 \, \rho_p$, depending on the flow regime (see \mbox{Equation~(\ref{eq:tauf})}). Thus, for~a fixed value of the mass density of the dust particle, the~Stokes number represents the particle size. Hence, particles with different sizes (or Stokes numbers) obtain different speeds at the same heliocentric distance, which leads to relative velocities among the solid particles and, consequently, to~collisions.
\par The frictional interaction of gas and dust and the subsequent sub-Keplerian speed of the dust particles result in an inward drift of the dust, with~radial velocities depending on Stokes number (i.e., particle size). Very small and very large objects possess radial velocities that are very small; the maximum radial speed is reached for dust particles with $St = 1$. In~general, dust drifts towards higher gas pressures, which also applies for vertical settling towards the disc midplane or local pressure bumps. 
\par Turbulent gas motion influences the dust motion as well. Quantitatively, the~turbulence of a disc can be described by the turbulent strength parameter $\alpha$. Qualitatively, gas turbulence can lead to collisions among the dust particles as well as to particle transport. Thus, thermally processed particles from the inner disc might be transported in- or outward due to turbulence, which leads to radial mixing of the dust species formed at different temperatures. 

\par For very small, i.e.,~sub-micrometre to micrometre-sized, grains, Brownian motion may also be responsible for very-low-speed collisions. Due to the Maxwell--Boltzmann distribution of the mean thermal kinetic energy of the grains, equal-sized particles can also collide due to Brownian motion~\cite{Weidenschilling.1980}.

\subsection{A Growth Scenario from Grains to Pebbles}
\label{sec:DustCollision_Phase1}
When dust or ice particles collide, the~outcome depends on material properties, particle size and collision velocity~\cite{Weidenschilling.1977,Dominik.1997,Blum.2000,Blum.2008,Wada.2008,Wada.2009,Guttler.2010,Schrapler.2018}. The~relative velocities between the grains, the~number of particles of a given size and the collisional cross-sections determine the number of collisions per unit time. The~collision velocities result from the processes that induce the relative motion between dust and gas (see Section \ref{sec:PPD}). For~radial, azimuthal and vertical drift, the~velocity between dust and gas depends on particle size; hence, only particles of different sizes collide, because~particles of equal size move with the same speed. However, equally sized particles can collide in case of stochastic motion, as~induced by Brownian motion or gas~turbulence. 

\par When particles of comparable size collide, the~outcome can be sticking, bouncing, abrasion or fragmentation. Sticking always occurs when the collision energy is small compared to the van der Waals binding energy of the formed contact~\cite{Dominik.1997}. In~collisions at higher impact energies, the~outcome is determined by the degree of inelasticity, which can be a function of temperature~\cite{Jankowski.2012,Gundlach.2015}. 

The hit-and-stick process can be accompanied by deformation and compaction~\cite{Guttler.2010}. However, in~the ultra-low velocity regime, compaction does not occur, and~the growing agglomerates develop a fractal structure~\cite{Dominik.1997,Kempf.1999,Blum.2000b,Krause.2004,Blum.2006}, with~a fractal dimension of \linebreak $D_f \lesssim 1.9$ \cite{Dominik.1997, Blum.2000b, Paszun.2009,Wada.2008}. For~higher collision speeds, the~fractal dimension may increase to $D_f \thickapprox 2.5$ \cite{Wada.2008} or even to the limit of $D_f \thickapprox 3.0$. High-speed collisions can lead to fragmentation of the agglomerates~\cite{Blum.1993,Beitz.2011,Schrapler.2011,Deckers.2013,BukhariSyed.2017}. The~mass ratio of the largest fragment to the initial mass decreases with increasing collision speed~\cite{BukhariSyed.2017}. In~the transition regime between sticking and fragmentation of similar-sized aggregates, bouncing occurs~\cite{Blum.1993,Heielmann.2007,Weidling.2012,Kothe.2013,Brisset.2016,Brisset.2017}. Bouncing collisions are inelastic and result in compaction. Successive bouncing events result in a volume-filling factor of the packing of $\sim$$0.36$ \cite{Weidling.2009}. We refer to these particles as porous~agglomerates. 

\textls[-15]{When a small aggregate (projectile) hits a larger one (target), along with the outcomes mentioned above, mass transfer, cratering and erosion are also possible. Mass can be transferred from the projectile to the target, while the projectile fragments~\cite{Wurm.2005b,Teiser.2009,Teiser.2009b,Guttler.2010,Teiser.2011,Beitz.2011,Meisner.2013,Deckers.2014,BukhariSyed.2017}. The~mass-transfer efficiency reaches from a few percent to roughly $50$ percent of the projectile’s mass, and~the mass-transfer process leads to compaction of the target, with typical volume-filling factors of $0.3$ to $0.4$ \cite{Teiser.2011,Meisner.2013}.} For~smaller size ratios between projectile and target, cratering occurs instead of net mass transfer. In~this case, more mass is lost in the form of ejecta than transferred  to the target~\cite{Wurm.2005,Paraskov.2007}. The~relative mass loss of the target agglomerate mainly depends on the impact energy~\cite{BukhariSyed.2017} and material strength, and~can be up to $35$ times the projectile mass~\cite{Wurm.2005,Paraskov.2007}. For~very small projectile-to-target size ratios, erosion (mass loss without a visible crater) has been experimentally observed for silica dust~\cite{Schrapler.2011,Schrapler.2018} and water ice~\cite{Gundlach.2015}, and~has also been numerically studied~\cite{Seizinger.2013,Krijt.2015}. The~efficiency of erosion depends on the impact velocity and the projectile mass~\cite{Schrapler.2018}.

In general, all mentioned processes depend on material and disc properties, e.g., monomer-grain material and size distribution, PPD model, turbulence strength and distance to the central star. Different materials can differ significantly in properties, such as sticking threshold or tensile strength, as~shown, e.g.,~for organic materials~\cite{Bischoff.2020}. This was also observed for collisions of “interstellar organic matter analogues”, which showed enhanced stickiness around $250 \,\mathrm{K}$, but~less sticking for lower or higher temperatures~\cite{Kouchi.2002,Kudo.2002}. For~$\mathrm{CO_2}$, the~sticking threshold seems to be comparable to that of silica~\cite{Musiolik.2016}.  Recently, Arakawa et al. 2021 \cite{Arakawa.2021c}  reconciled the different experimental results on the sticking--bouncing threshold for water ice, carbon-dioxide ice and silica particles. Besides~the surface energy (closely related to the tensile strength) of the material, the~visco--elastic dissipation of energy is the essential material property, and~is much larger for water ice at low temperature than for the other two materials. Hence, water ice sticks at higher velocities than carbon dioxide or silica particles of the same size, in~agreement with laboratory results~\cite{Gundlach.2015}.

\textls[-20]{Several barriers prevent the further collisional growth of dust agglomerates to kilometre-sized planetesimals. The~most important ones will be briefly described~below.}

First, drift of particles into the central star, caused by the friction between dust and gas, is a limiting factor. The~drift efficiency depends on particle and disc properties, with~a drift time scale of
\begin{equation}
   t_d = a/v_d = a/(2 \, \mathrm{St} \, v_\mathrm{max}).
   \label{eq:raddrift}
\end{equation}

Here, $a$, $v_d$, $\mathrm{St}$ and $v_\mathrm{max}$ are the heliocentric distance, radial drift velocity, Stokes number of the dust particle and maximum deviation of the orbital gas velocity from Keplerian velocity, respectively. Equation~(\ref{eq:raddrift}) is valid for $\mathrm{St} \leq 1$, and~typical values for $v_\mathrm{max}$ are \mbox{$50 \,\mathrm{m/s}$
\cite{Weidenschilling.1977})}. This barrier provides a critical time constraint for the formation of planetesimals. The~maximum radial drift velocity of $2 \, v_\mathrm{max}$ is reached for bodies with $\mathrm{St}=1$, which means such particles possess drift timescales on the order of $t_d \approx 100 \, \mathrm{yr}$ for $a=1 \, \mathrm{au}$. This timescale is usually much shorter than the collisional growth time beyond $\mathrm{St}=1$.

Second, the~so-called bouncing barrier stops growth when the sticking--bouncing threshold is reached and the colliding aggregates bounce off. Due to the inelasticity of the bouncing process, compaction of the aggregates occurs~\cite{Weidling.2009,Zsom.2010,Lorek.2018}. The~bouncing barrier is reached at aggregate sizes of millimetres to centimetres, depending on the particle and disc properties~\cite{Zsom.2010,Lorek.2018}. The~resulting agglomerates have been termed “pebbles”. The~largest pebbles are achieved in the minimum mass solar nebula model~\cite{Zsom.2010}. For~increasing heliocentric distance or increasing dust-to-ice ratio, the~maximum pebble size decreases and the pebbles become more porous~\cite{Lorek.2018}. Millimetre- to centimetre-sized solid particles have been observed in PPDs~\cite{vanBoekel.2004,DAlessio.2006,Birnstiel.2010,Ricci.2010,Perez.2012,Trotta.2013,Testi.2014,Perez.2015,Tazzari.2016,Liu.2017}.

Third, even if the bouncing barrier can be overcome, further growth is halted because collisions result in destruction of the aggregates. Fragmentation typically happens at collision speeds of $\gtrsim$1 {m/s} \cite{Blum.2008}.

Fourth, recently, the~erosion barrier was experimentally discovered~\cite{Schrapler.2018}. Erosion is present when dust grains, or~small aggregates thereof, impinge larger dust aggregates at high speeds. In~this case, impacts liberate other grains from the aggregate surfaces so that the mass of the aggregates is slightly reduced. Numerical simulations have shown that this erosion is a runaway effect even if all other other collisions are assumed to result in sticking. Due to the erosion barrier, maximum aggregate sizes are on the order of $0.1 \, \mathrm{m}$ \cite{Schrapler.2018}.

Particles in the weakly ionized disc can carry a nonzero negative charge, which also influences the collisional behaviour. Due to the asymmetric distribution of charges, a~repulsive force acts between the particles, hindering sticking~\cite{Okuzumi.2009,Okuzumi.2011,Okuzumi.2011b}. This effect is called the electrostatic barrier. It can already halt growth at sizes of fractal agglomerates~\cite{Okuzumi.2011}. When taking the photoelectric effect into account, a~layer of uncharged dust can build up, which might overcome the electrostatic barrier~\cite{Akimkin.2015}. However, in~low turbulence regions of the disc, such as the dead zone, growth can be effectively suppressed~\cite{Akimkin.2020}. Numerical models have also shown that the charge of the particles influences growth rate, size and compactness, as~highly charged particles grow to larger sizes and get more compact~\cite{Xiang.2020}, which influences the bouncing and fragmentation thresholds. This may lead to a bridge between bouncing barrier and pebble sizes required for streaming instability, as~indicated by experimental and numerical work~\cite{Steinpilz.2020} and could therefore also be beneficial for growth. However, the~electrical charging of particles and its influence on the growth process need to be studied further to understand the whole~picture.

To summarise, the~formation of pebbles, millimetre- to centimetre-sized agglomerates of microscopic dust and ice grains, can be understood with current empirical knowledge about the collision behaviour of protoplanetary dust. However, any further growth seems impeded by the presence of a number of barriers, which --- at the current stage of research --- makes the direct growth of planetesimals impossible. Thus, other processes that lead to the formation of planetesimals are required. In~the next Section, we show how this might be~possible.

\section{Collapse of Pebble Clouds into Planetesimals}
\label{sec:Formation_of_Planetesimals}
\subsection{Pre-concentration of Pebbles}
\label{sec:Preconcentration}

Regions of enhanced concentration of pebbles can be formed in eddies, vortices and pressure bumps of turbulent PPDs~\cite{Johansen.2014}. Vortices and pressure bumps can be caused by several phenomena, e.g.,~magneto-rotational instability~\cite{Balbus.1991}, baroclinic instabilities~\cite{Klahr.2003,Lyra.2011}, tidal forces of early formed planets and gaps induced by them~\cite{Lyra.2009}, the~edges of the dead zone~\cite{Lyra.2008} and snow lines (e.g., \cite{Dzyurkevich.2010}). For~example, in~the transitional disc IRS 48, large-scale vortex structures with a  concentration of millimetre-sized pebbles were observed~\cite{vanderMarel.2013}. 

Turbulent rotating structures can be described by their spatial dimension $l$, their rotation velocity $v_e$ and their turnover time $t_e = l/v_e$. For~example, particles with stopping times of $\tau_f \thickapprox t_e$ are efficiently concentrated in high-pressure regions between eddies, which results in a narrow size range for such concentrated particles. Coriolis force and shear are dominant on larger timescales. In~such cases, particles with $\tau_f \sim (2 \Omega)^{-1} \ll t_e$ are efficiently trapped, independent of the eddy turnover time. When also taking Keplerian shear into account, the~optimal stopping time is $\tau_f = \Omega^{-1}$. The~radial drift speed can then be deduced from the eddy velocity, and~reads
\begin{equation}
    v_r = - \frac{2 \Delta v}{St^{-1} + St} \, .
\end{equation}
Here, the~local value of $\Delta v$ is important. For~example, if~$\Delta v = 0$, i.e.,~in pressure bumps, the~particles are being concentrated~\cite{Johansen.2014}.

\subsection{Further Concentration of Pebbles by the Streaming~Instability}
\label{sec:SI}
Due to the concentration of particles as described above, the~collective surface-to-mass ratio of a pebble cloud is reduced and consequently also the gas drag. If~the local dust-to-gas mass ratio exceeds unity, streaming instability is triggered and concentrates the pebbles further~\cite{Youdin.2005}. The~velocity of the pebble cloud reaches Keplerian velocity, and~radial drift is reduced. As~a back-reaction by the pebble ensemble, the~gas is forced to move at Keplerian velocity. Individual pebbles or small pebble clusters on the same orbit, which still move at reduced speed due to gas friction, are overtaken by the pebble cloud. Inward-drifting pebbles are also incorporated into the pebble ensemble when crossing the respective orbit. Thus, the~mass of the pebble cloud rapidly grows over a short timescale. An~upper limit of the growth timescale was found to be $\sim$$10^5$ years~\cite{Carrera.2015}. The~length scale of streaming instability is typically 5 percent of the gas scale height~\cite{Yang.2014}, and~a dust concentration of up to several thousands can be reached~\cite{Bai.2010, Johansen.2012}. The~threshold when streaming instability is triggered can also be expressed by the metallicity $Z = \sum_{peb}/\sum_{g}$, i.e.,~the surface density ratio of pebbles and gas. For~Stokes numbers $St \thickapprox 0.1$ (which corresponds to pebble sizes of  $\sim$0.1 $\,\mathrm{m}$ at $1 \,\mathrm{AU}$ and $\sim$$1 \,\mathrm{mm}$ at $100 \,\mathrm{AU}$), a~minimum metallicity of $Z_\mathrm{min} \thickapprox 0.015$ is required, and~$Z_\mathrm{min}$ increases for larger and smaller Stokes numbers~\cite{Youdin.2005,Carrera.2017}. As~the solar metallicity is $Z = 0.0134$ \cite{Asplund.2009}, enhancement in $Z$ by partial dissipation of nebula gas might be required (see discussion in~\cite{Davidsson.2016}). However, recent results show the possibility for particle clumping due to streaming instability for smaller metallicities~\cite{Li.2021b}.
Substructures of the concentrated regions have been observed in computer simulations~\cite{Johansen.2012}. The~resilience of streaming instability was investigated by Carrera et al. 2021 \cite{Carrera.2021b}, who found it to be very robust for centimetre-sized pebbles and in the case of only weak pressure~bumps. 

Most studies assume monodispersed pebble sizes. However, this is certainly an academic assumption. Recent works on streaming instability with polydispersed pebble sizes show that the size--frequency distribution has an important influence on, e.g.,~the parameters for instability and growth time scales~\cite{Paardekooper.2020,Paardekooper.2021,McNally.2021,Zhu.2020,Schaffer.2021}. Additionally, including turbulence reduces the growth rates compared to laminar discs~\cite{Umurhan.2020} and increases the critical metallicity, depending on Stokes number, needed to trigger streaming instability~\cite{Carrera.2017}.

\subsection{Gravitational Collapse}
\label{sec:GravitationalCollapse}
If the mass concentration of the pebble cloud reaches the Roche density
\begin{equation}
    \rho_R = \frac{9 \Omega^2}{4 \pi G} 
\end{equation}
\textls[-25]{with gravitational constant $G$, gravitational collapse occurs, as~described by Johansen et al. 2014 \cite{Johansen.2014}. }

The cloud collapses into a gravitationally bound body. Whether the pebbles survive the collapse depends on the mass of the pebble cloud. Pebbles can be destroyed either during collisions in the free-fall phase or upon impact on the surface of the forming planetesimal~\cite{BukhariSyed.2017,WahlbergJansson.2017b}. Moreover, the~hydrostatic pressures inside the planetesimal can also destroy the pebbles~\cite{Blum.2018}. Thus, the~integrity of the pebbles is only guaranteed for low-mass planetesimals. Estimates on the strength of pebbles have shown that this criterion is satisfied for planetesimals with radii $\lesssim 50 \, \mathrm{km}$. Size sorting during the collapse could lead to stratification of the pebble size inside a planetesimal~\cite{WahlbergJansson.2017}. It has been argued that during the collapse, the~pebbles may capture smaller dust/ice particles, including those still in the fractal growth stage~\cite{Fulle.2017}. If~the pebbles survive the collapse intact, the~pore spaces between the pebbles may not be entirely~empty.

The radii of planetesimals formed by streaming instability with gravitational collapse are predicted to be between 50 and 1000 km, depending on the place of birth in the PPD and the disc properties~\cite{Johansen.2007,Johansen.2009,Johansen.2011,Johansen.2012,Kato.2012, Gerbig.2020,Gole.2020, Klahr.2020b}.  With~increasing disc mass, the~formed planetesimal sizes increase~\cite{Johansen.2012}. The~resulting mass--frequency distribution function of planetesimals can be fitted with a power law with exponent $-1.6 \pm 0.1$ \cite{Simon.2016, Simon.2017,Schafer.2017}, but~it was also found that an exponentially truncated power law with a low-mass slope of $-1.3$ fits better~\cite{Abod.2019}. Klahr and Schreiber 2020 \cite{Klahr.2020b} discuss that a Gaussian distribution of the initial planetesimal size could provide a reasonable fit. High variability of the resulting planetesimal size distribution for simulations with slightly different initial conditions was found by Rucska et al. 2020 \cite{Rucska.2020}. 

However, due to restrictions of the numerical simulation resolution, the~lower threshold of the planetesimal size is unknown. With~increasing numerical resolution, ever smaller pebble clouds and their collapse can be resolved~\cite{Johansen.2012}. Therefore, it is an open question as to which sizes of planetesimals can be formed by streaming instability and the subsequent gravitational collapse. Klahr and Schreiber 2020 \cite{Klahr.2020b} suggest that large planetesimals form first and closer to the Sun, and~the formation of smaller planetesimals is possible further out in the PPD. Consequently, these objects would form later in~time. 

It seems likely that pebble clouds can also collapse into binaries, which can merge into contact binaries via a low-velocity impact~\cite{Nesvorny.2010,Nesvorny.2019}. 
The influence of the water-ice snowline opens the possibility for a bifurcation in planetesimals with~different properties inside and outside the snowline~\cite{Lichtenberg.2021}. 

As stated above, planetesimals larger than $50 \,\mathrm{km}$ in radius are mechanically altered under their own weight. Pebbles are compacted and destroyed through local hydrostatic pressure, depending on their radial position inside and the total mass of the planetesimal. The~loss of the pebble structure is accompanied by an increase of the tensile strength from $\sim$$1 \, \mathrm{Pa}$ \cite{Blum.2014} to $\sim$1--10 $\, \mathrm{kPa}$ \cite{Blum.2006} due to considerable increase in grain--grain contacts when the pebbles are compressed. Thus, cometary activity by gas pressure cannot be explained without pebbles, because~pressures in excess of a few Pa cannot be reached in the cometary subsurface regions~\cite{Grundy.2020}. Because~we focus on comet formation in this review, we further exclude planetesimals larger than $50 \,\mathrm{km}$ in~radius.

\section{Evolution of Planetesimals towards Comets}
\label{sec:Evolution}
In this Section, we discuss the main evolutionary processes acting on the bulk of the planetesimal volume over a timespan of 4.6 Gyrs. This encompasses, on~the one hand, internal heating from radioactive decay of short-lived and long-lived radionuclei, and, on~the other hand, possible collisional encounters of the planetesimal with other bodies in the outer regions of the solar~system.
\subsection{Radiogenic Heating}
\label{sec:Radiogenic Heating}

Due to the decay of radioactive isotopes, planetesimals can be heated in their interior. This alters the body dramatically if critical temperatures are reached. In~this case, sublimation of volatiles depletes ices in the central regions, where temperatures are typically higher than those closer to the surface. Consequently, the~advecting volatiles recondense in cooler regions and thus accumulate in shells in the outer parts of the planetesimals. At~higher temperatures and pressures, some materials can also melt. Both melting and vaporisation lead to differentiation of the objects. The~efficiency of radiogenic heating mainly depends on the size of the planetesimal, the~thermal conductivity of the material and on the amount and half-lives of the radiogenic isotopes incorporated into the~body.

The radioactive materials are divided into two groups, short-lived radionuclides (SLNs) and long-lived radionuclides (LLNs).  SLNs (see Section~\ref{sec:FirstRadiogenicHeating}) possess half-lives typically on the order of one to a few million years, which means that their heating effectively occurs only early in the lifetime of the solar system in the PPD phase. Moreover, as~SLNs decay quickly, the~formation time of the planetesimals also matters. In~contrast, LLNs decay over billions of years (see Section~\ref{sec:SecondRadiogenicHeating}).

Along with thermal alteration due to radiogenic heating, solar illumination can also increase the temperature, depending on long-term orbital dynamics. As~shown by modelling of Jupiter-family comets on their path inwards from the scattered disc or the Kuiper Belt~\cite{Gkotsinas.2022}, this external thermal alteration can penetrate deep into the body, so that hypervolatiles may be redistributed and lost completely. Further, temperatures above 80 or even 110 K can be reached; hence, super-volatiles and amorphous water ice can be affected, too. 

\subsubsection{Short-lived Radionuclides}
\label{sec:FirstRadiogenicHeating}
The effect of SLNs is dominated by two nuclei, namely $^{26}\mathrm{Al}$ and $^{60}\mathrm{Fe}$. It seems that the distribution of $^{26}\mathrm{Al}$ was homogeneous in the solar nebula with a ratio of $^{26}\mathrm{Al}(t=0)/^{27}\mathrm{Al} = 5 \times 10^{-5}$ \cite{Gregory.2020}, with~$t=0$ denoting the time of the formation of calcium--aluminium-rich inclusions (CAIs). Due to their short half-lives, SLNs can heat bodies to high temperatures. Even the melting point of water ice can be reached in planetesimal-sized objects~\cite{Prialnik.1995}. In~the last two decades, more and more complex studies have been performed to understand the influence of radiogenic heating on the evolution of planetesimals~\cite{Choi.2002,Merk.2002,Merk.2003,Merk.2006,Sanctis.2007,Mousis.2012,Holm.2015,Mousis.2017,Golabek.2021,Lichtenberg.2021}. Most of these studies neglected LLN and did not consider pebble-structured objects. The~sizes of bodies investigated in their thermal evolution vary from a few kilometres to several $100 \,\mathrm{km}$ in radius. In~addition, formation time has a significant influence on the evolution of the objects; for example, Lichtenberg et al. 2021 \cite{Lichtenberg.2021} focused on early formed bodies with formation times $< 3 \,\mathrm{Myrs}$ and found mean temperatures of $\sim$$150 \,\mathrm{K}$, even for bodies as small as $1 \,\mathrm{km}$ in radius. Mousis et al. 2017 \cite{Mousis.2017} found for planetesimals with radii of $1.3 \,\mathrm{km}$ and $35 \,\mathrm{km}$ that they need to be formed later than $2.5 \,\mathrm{Myrs}$ and $5.5 \,\mathrm{Myrs}$ after CAI, respectively, to~prevent amorphous water ice from crystallising. Golabek and Jutzi 2021 \cite{Golabek.2021} include, along with radiogenic heating, heating by collisions, and~provide estimates for the maximum radius and formation time of planetesimals to avoid heating above $40 \,\mathrm{K}$ in the interior. The~maximum radius can be up to $40 \,\mathrm{km}$ for a formation later than $5 \,\mathrm{Myrs}$ after CAI, and~even $100 \,\mathrm{km}$, including catastrophic collisions for a critical temperature of $80 \,\mathrm{K}$ (see Figure~9 in Golabek and Jutzi 2021 \cite{Golabek.2021}).  

However, the~thermal conductivity of the body, which is linked to the internal structure and composition, is of major importance for modelling radiogenic heating. In~cases of high thermal conductivity, the~heat can be transported efficiently to the surface, where it can be radiated away, resulting in lower peak temperatures, whereas low thermal conductivities impede energy transport from the interior towards the outer regions. Consequently, the~peak central temperature can be very~high. 

It is the pebble-pile structure that makes planetesimals formed by gentle gravitational collapse so interesting for the study of their thermal evolution, because~(i) pebbles possess low thermal conductivity due to their porous structure, (ii) the thermal contact between pebbles is minimal, and~(iii) radiative heat transport is very inefficient at low temperatures. The~heat-transport model of Gundlach et al. 2012 \cite{Gundlach.2012} shows that at low temperatures and for pebbles with radii of $5 \, \mathrm{mm}$, the~thermal conductivity is less than $10^{-3} \, \mathrm{W \, m^{-1} \, K^{-1}}$ (see also Figure~1 in~\cite{Bischoff.2021}) and, thus, is more than three orders of magnitude lower than that of solid (i.e., non-porous, non-hierarchical) materials with the same bulk~composition.

The study by Malamud et al. 2022 \cite{Malamud.2022} has for the first time included the pebble structure of small planetesimals (see Section \ref{sec:GravitationalCollapse}) in their analysis of radiogenic heating by SLNs and LLNs. They found that even for planetesimals as small as $0.5$ to $20 \,\mathrm{km}$ in radius, the~influence of radioactive heating can be significant. Only bodies up to $2 \,\mathrm{km}$ in radius with a late formation time of $>$5$ \,\mathrm{Myrs}$ after CAI obtain a peak temperature below $70 \,\mathrm{K}$. Higher temperatures force the (super-)volatiles to diffuse outwards and to condense in regions where the temperatures are below the condensation temperature. Differentiation of the ices, with~a depleted interior and an enriched outer layer, is the result of this process. However, the~near-surface regions are unaltered because outward diffusion stops below the surface. This differentiation of the volatiles is visualised in Figure~\ref{fig:overview} with red (pebbles depleted in ices) and white (ice-enriched shell) colours. When water ice is redistributed within the planetesimal, no hyper- or super-volatiles can remain in the water-ice depleted interior. Reaching the melting point of water ice was observed in the model when planetesimal radii are $5 \,\mathrm{km}$ or larger; these planetesimals were formed rapidly after CAI and have large mineral~abundance.  

The evolved planetesimals after the phase of radiogenic heating by SLNs can be divided into undifferentiated and differentiated objects. Small planetesimals that avoided excessive heating remain undifferentiated and are therefore considered pristine (\mbox{Type A}). The~differentiated planetesimals (Type B) are split into subgroups. The~first stage of differentiation is the redistribution of super-volatiles, as~observed in the simulations of Malamud et al. 2022 \cite{Malamud.2022}. The~most crucial differentiation occurs when melting temperatures and pressures (e.g., of~water) are reached. In~this case, aqueous alteration starts and pebble structures are destroyed, which also decreases porosity. In~general, radiogenic heating by SLNs seems to be an important evolutionary process for most planetesimals, except~for the smallest ones, which were formed late after CAI. Within~the timescale of the first radiogenic heating phase of up to $10 \,\mathrm{Myrs}$ after CAI, the~influence of collisions can often be~neglected.

\subsubsection{Long-lived Radionuclides}
\label{sec:SecondRadiogenicHeating}
In contrast to the heating by SLNs, LLNs decay over a much longer period of time. Relevant LLNs for planetesimals are $^{235}\mathrm{U}$, $^{238}\mathrm{U}$, $^{40}\mathrm{K}$ and $^{232}\mathrm{Th}$, and~these materials are important for larger objects and for long time scales. Their influence was first analysed by Whipple 1965 \cite{Whipple.1965} and Prialnik et al. 1987 \cite{Prialnik.1987}. Both publications found that LLNs can increase the central temperature by a few $10 \,\mathrm{K}$. The~second phase of radiogenic heating is accompanied by collisional evolution (see Section \ref{sec:CollisionalEvolution}).

\subsection{Collisional Evolution}
\label{sec:CollisionalEvolution}
Besides internal heating, collisions among planetesimals can have a significant influence on their evolution. Collisions can occur over the whole time-range after formation until today. However, collision probability evolves over time and depends on the planetesimal size and orbit. Hence, bodies with different sizes and thermal evolutions may collide with one another. The~outcome of these collisions depends on the physical properties of the planetesimals, the~collision angle, the~impact velocity and the size of the colliding bodies. In~this review, we only consider collisions that have potentially global impact on the planetesimals, and~define three possible collisional types, defined~below. 
\begin{enumerate}
\item First, a~collision can be \emph{sub-catastrophic}  
  if the largest fragment possesses at least half of the initial mass of the larger bodies. We assume that in this case, the~mechanical bulk properties of the planetesimals are essentially conserved.
\item Second, if~the fragments are smaller, the~collision is considered \emph{catastrophic}. 
 However, if~the energy input into the colliding bodies was not sufficient to destroy the pebble structure, the~fragments are still composed of pebbles, but~the original bulk properties might have changed.
\item Third, the~collisions can be \emph{super-catastrophic}, in~which case the pebble structure and the pebbles themselves are destroyed by the collision event.
\end{enumerate}

Several researchers have performed simulations of the collisions of planetesimals for different sizes and impact velocities. Abedin et al. 2021 \cite{Abedin.2021} investigated the collision probabilities in the trans-Neptunian region of today and found a wide range of possible radii and velocities of colliding bodies, but~favouring low relative speeds ($< 1 \,\mathrm{km/s}$) and the main classical belt ($30 \,\mathrm{AU} < R_h < 50 \,\mathrm{AU}$) as the collision site. The~scattered and detached populations of the trans-Neptunian region show the smallest collision probability. The~physical parameters are connected to those introduced in Section~\ref{sec:Coagulation} of the dust and ice particles, and~the pebbles thereof, and~supported by experimental work. However, for~planetesimal-sized objects, experimental results need to be extrapolated from laboratory scales. In~addition, the~impact angle can have an important influence on the induced heating~\cite{Davison.2014}. However, impact heating is most relevant for impact velocities of several kilometres per second and can cause melting of the material. It is interesting to note that porous bodies are heated more efficiently by collisions~\cite{Davison.2010} than solid objects. The~numerical study of Charnoz et al. 2007 \cite{Charnoz.2007} found that the formation of the Kuiper belt, the~scattered disc and the Oort cloud can be explained by dynamical depletion of those regions or low-efficiency implantation of bodies, in~both cases accompanied by little collisional~activity. 

In the following subsections, we briefly review the previous work of sub-catastrophic, catastrophic and super-catastrophic collisions in the framework of comet~formation.

\subsubsection{Sub-Catastrophic Collisions}
\label{sec:Collision_Sub_catastrophic}
We term the cases in which the largest fragment after a collision has at least half of the initial mass of the larger of the two colliding bodies as sub-catastrophic collisions. These include cratering, erosion and collisions resulting in mass growth, such as merging and mass transfer. Golabek and Jutzi 2021 \cite{Golabek.2021} studied the combined effect of radiogenic heating and collisions and found that only the unbound material is heated significantly, whereas the material that remains bound is hardly effected. A~study of Jutzi and Asphaug 2015 \cite{Jutzi.2015b} showed the possibility of the formation of a bilobate shape by a low-velocity collision, assuming non or small cohesion ($\leq$$100 \,\mathrm{Pa}$) and a collision speed of $1.5 \,\mathrm{m/s}$. Furthermore, the~formation of bilobed objects by sub-catastrophic collisions was shown by Jutzi and Benz 2017 \cite{Jutzi.2017}, who derived that such an evolution is more likely to happen than collision-avoiding or a catastrophic breakup of planetesimals. In~the following, we assume that sub-catastrophic collisions do not change the morphology, micro- and macro-porosity, mechanical properties and volatile abundances of the largest surviving fragment. Thus, this fragment has the exact same properties as the planetesimal before the collision, with~the exception of mass and~shape.

\subsubsection{Catastrophic Collisions}
\label{sec:Collision_Catastrophic}
If the largest fragment after a collision possesses less than half of the initial mass, we term the collision as catastrophic. The~difference from a super-catastrophic collision (see Section \ref{sec:Collision_Super_catastrophic}) is that catastrophic collisions are not energetic enough to compact or destroy pebbles or to significantly reduce the amount of volatiles. The~formation of comet 67P from a catastrophic collision and the following re-accumulation of its current mass, with~conservation of volatiles and low bulk density, was proposed by \mbox{Schwartz et al. 2018 \cite{Schwartz.2018}}. Jutzi and Michel 2020 \cite{Jutzi.2020} showed that the material ejected and not re-accreted is indeed the most processed part of the original object. However, the~available simulations do not take a pebble structure of the colliding planetesimals into account. As~seen for radiogenic heating (see Section \ref{sec:Radiogenic Heating}), the~microphysical properties of the material can have a huge impact on the collisional outcome (see, e.g.,~\cite{Krivov.2018}). In~addition, it should also be studied how differentiated bodies behave in collisions. With~all these uncertainties in mind, we assume here that bodies that re-accumulated after a catastrophic impact still consist of pebbles and contain a large percentage of the volatiles contained in the planetesimal predecessors. In~contrast to the objects stemming from sub-catastrophic collisions (see Section \ref{sec:Collision_Sub_catastrophic}), bodies arising from catastrophic collisions are rubble piles and may consist of spatially separated parts stemming from different regions of the precursor planetesimals. We also assume that this rubble is so mechanically weak~\cite{Krivov.2018} that its outer shape does not survive the gravitational re-accretion process. Thus, re-accumulated bodies from sub-catastrophic collisions should not possess voids on length scales exceeding that of the pebble~size.

\subsubsection{Super-Catastrophic Collisions}
\label{sec:Collision_Super_catastrophic}
The term super-catastrophic collisions describes impacts that provide enough energy to destroy the pebble structure and to sublimate the volatile constituents. In~this case, no pebbles or ices will remain inside the object, which is indicated in Figure~\ref{fig:overview} by grey colour. Super-catastrophic collisions occur at high relative velocities. Objects that gravitationally re-accumulate after a super-catastrophic collision among planetesimals are rubble piles. In~contrast to the bodies stemming from catastrophic collisions (see Section \ref{sec:Collision_Catastrophic}), the~rubble from super-catastrophic collisions is highly compacted, and~thus mechanically strong enough to survive the accretion process intact. Thus, rubble piles from super-catastrophic collisions possess macro-porosity on length scales characteristic for the collisional outcome, and~reduced microporosity due to compaction of the dusty material upon~impact.

\section{Discussion - Which Planetesimal Can Become a Comet?}
\label{sec:Discussion}
With the limited knowledge about actual collisional encounters and outcomes (see Section \ref{sec:CollisionalEvolution}), we try in the following to physically characterise a planetesimal after 4.6 Gyrs of thermal and collisional evolution. The~formation and evolution scenarios described above may end in a variety of bodies, which we divide into three categories (see Figure~\ref{fig:overview} for an overview):
\begin{itemize}
    \item \textit{Icy Pebble Piles}---Type A1 and Type B1
    \item \textit{Icy Rubble/Pebble Piles}---Type A2 and Type B2
    \item \textit{Non-Icy Rubble Piles}---Type A/B3
\end{itemize}

One very important question for researching comets is: Which of these types could represent the comets currently observed in the solar system? In the following, the~most important physical properties, which can be used for the validation of the formation/evolution model and the corresponding observables, will be discussed. The~expectation from the formation and evolution scenario for each observable will be described and compared to real physical properties of comets measured by (spacecraft) observations. 

\subsection{Intrinsic Physical Properties}
\label{sec:Discussion:Intrinsic_Physical_Properties}
We will start this discussion with those physical properties that are directly provided by the formation and evolution process. These properties, such as internal morphology and total porosity, are referred to as intrinsic physical properties and can be used to directly link the formation and evolution of planetesimals with the observable properties of contemporary comets in the solar system. Table~\ref{tab:overview_body_types_intrinsic_properties} summarises the different intrinsic properties of cometary~nuclei.

\begin{table}[]
\caption{This table presents an overview of the different proposed types of objects with their expected intrinsic physical properties that are discussed in detail in the following sub-sections.\label{tab:overview_body_types_intrinsic_properties}}
		\newcolumntype{C}{>{\centering\arraybackslash}X}
		\begin{tabularx}{\textwidth}{CCCCCC}
			\toprule
				& \textbf{Type A1}	& \textbf{Type B1}     & \textbf{Type A2} & \textbf{Type B2} & \textbf{Type A/B3}\\
			 &   \includegraphics[width=1cm]{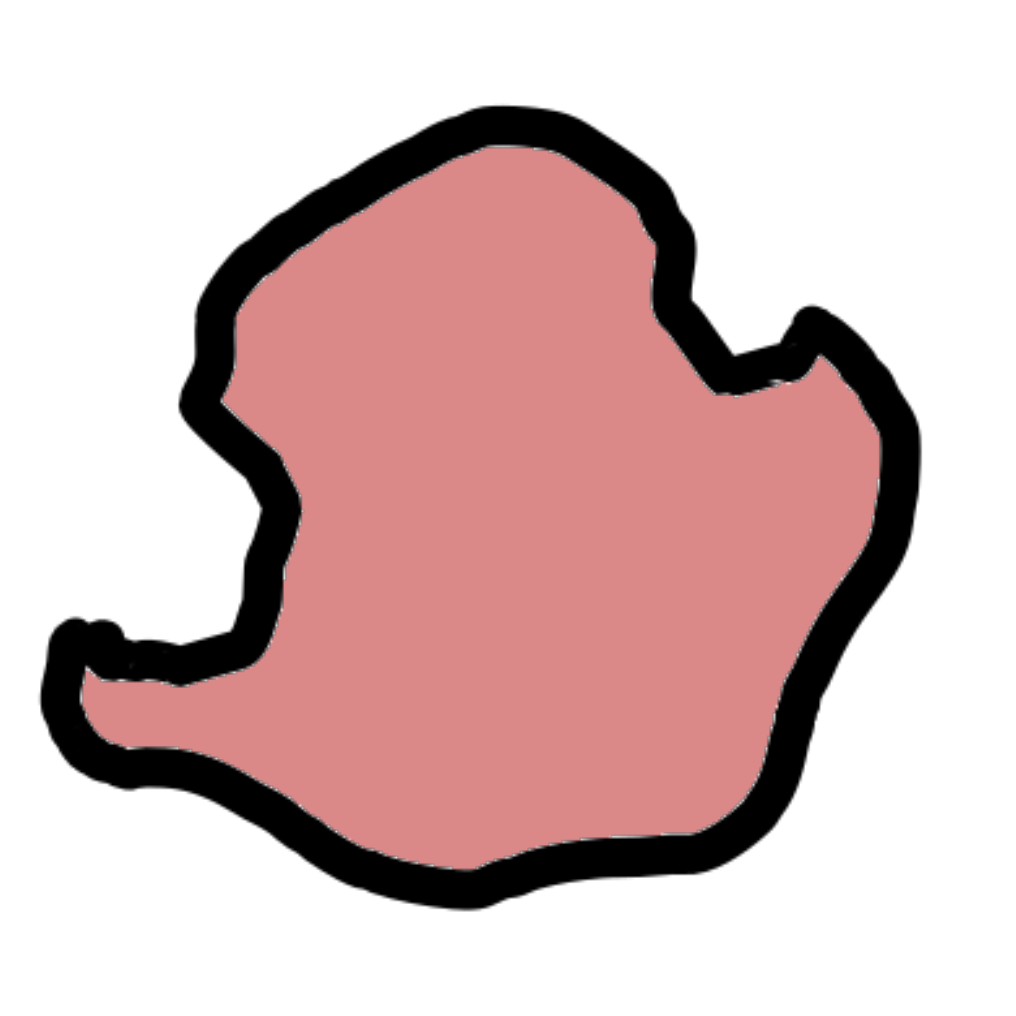} &   \includegraphics[width=1cm]{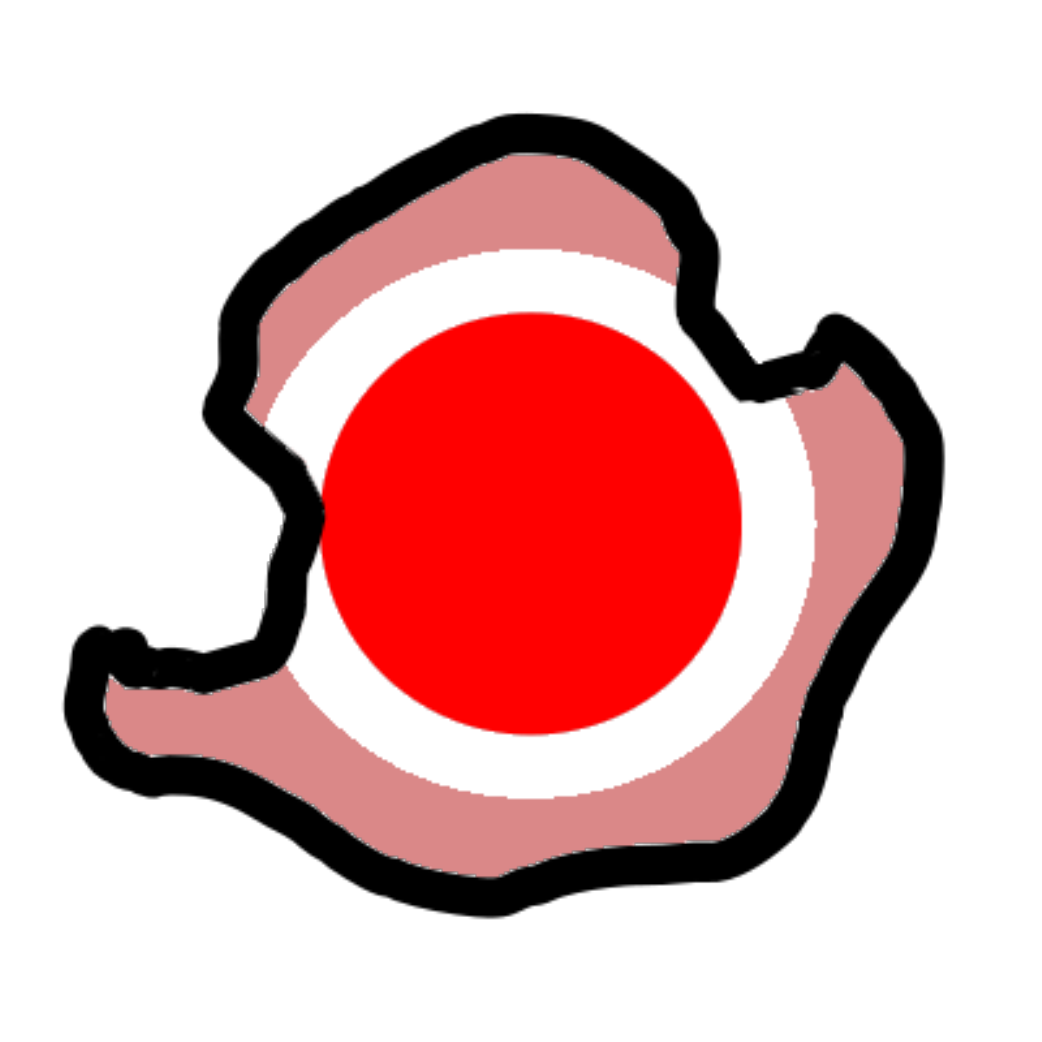} &   \includegraphics[width=1cm]{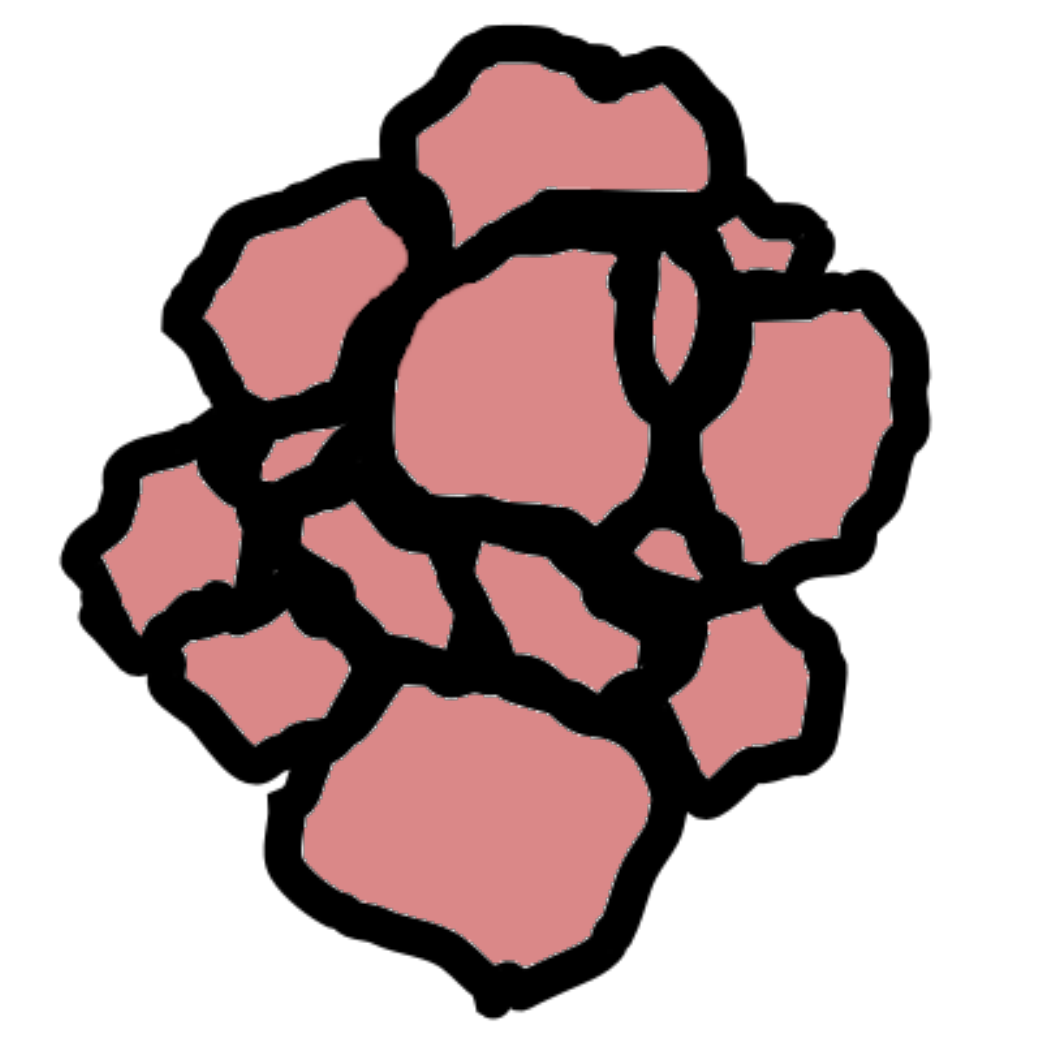} &  \includegraphics[width=1cm]{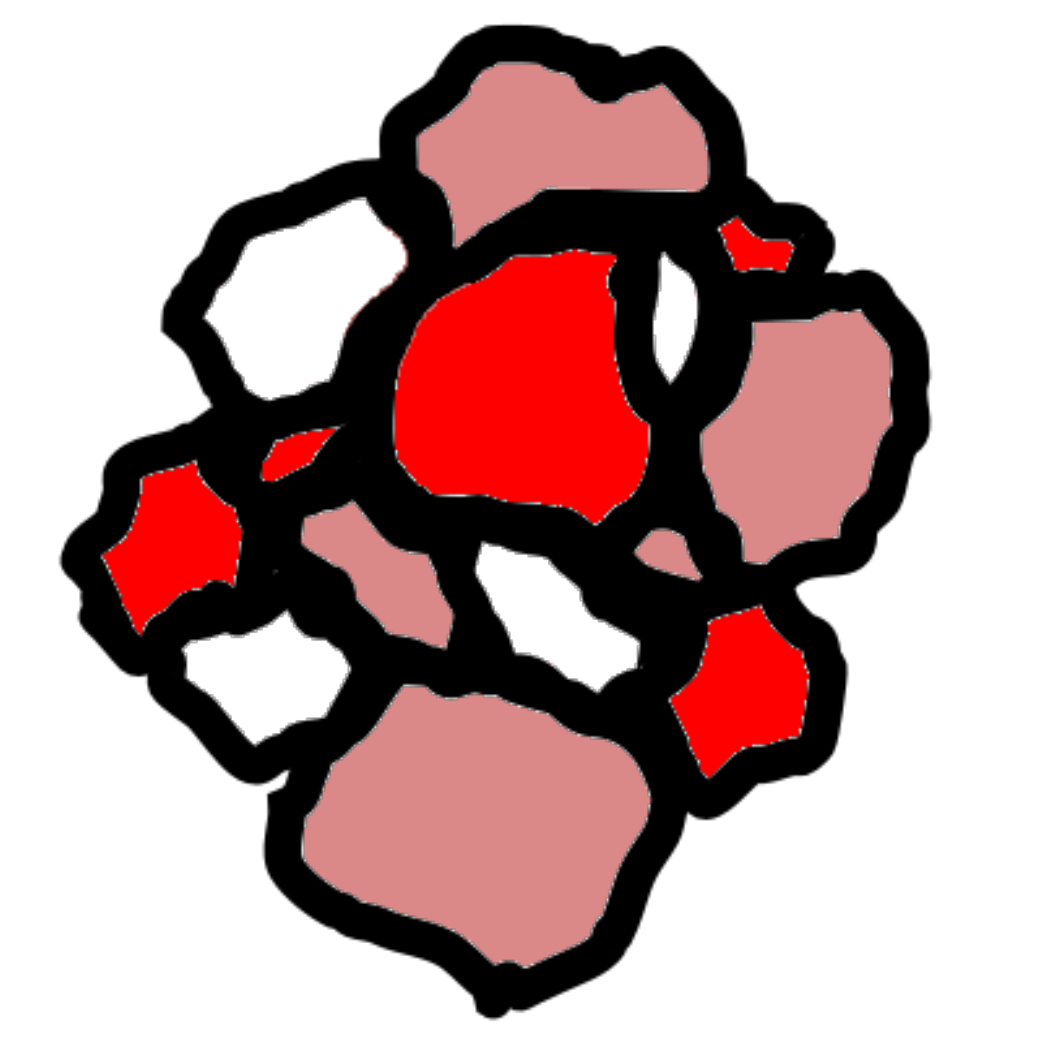} &   \includegraphics[width=1cm]{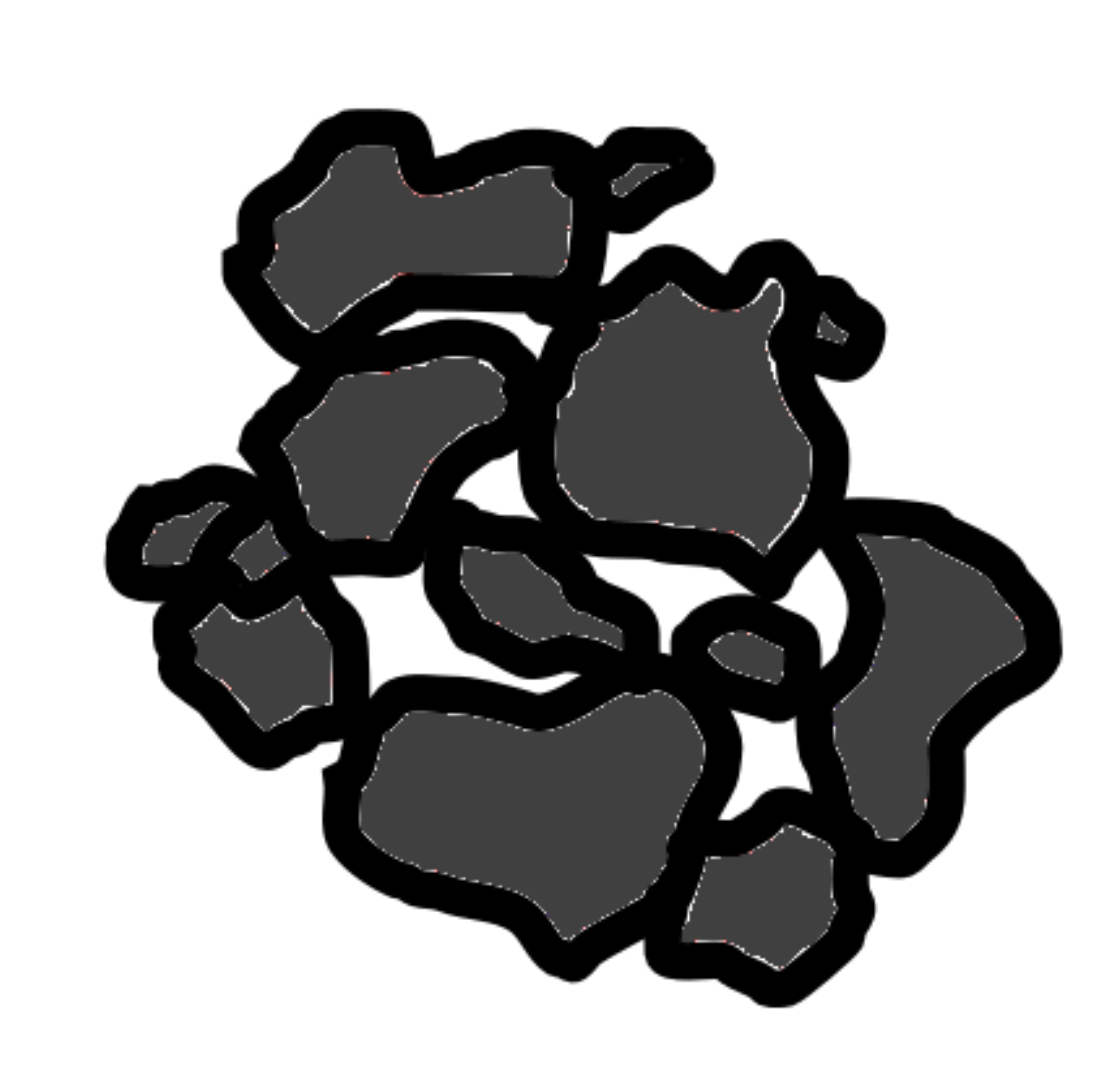} \\
			 \textbf{Intrinsic physical properties} & \multicolumn{2}{c}{Icy pebble piles} & \multicolumn{2}{c}{Icy rubble/pebble piles} & Non-icy rubble piles \\
			\midrule
            \textbf{Internal morphology and pebble properties (Section \ref{sec:Observables:InternalMorhology})} & Pristine, pebbles, void size equals pebble size, voids filled with fractals & Pristine, pebbles, void size equals pebble size, voids filled with fractals & Pebbles, void size equals pebble size, fractals locally depleted & Pebbles, void size equals pebble size, fractals locally depleted & Pebbles are destroyed and highly compacted, rubbles remain, large void spaces between rubbles \\
            \midrule
            \textbf{Total porosity (Section \ref{sec:Observables:TotalPorosity})} & High pristine porosity & Enhanced porosity in the interior, reduced porosity at ice shell locations, bulk porosity preserved & High pristine porosity & Heterogeneous, macro-scale mixture of material with enhanced and reduced porosity, pristine porosity may not be preserved & No intra-rubble porosity, macro-scale voids between rubbles, bulk porosity smaller compared to pristine porosity \\
            \midrule
            \textbf{Dust-to-ice mass ratio (Section \ref{sec:Observables:DIR})} & Pristine dust-to-ice ratio, homogeneous distribution & Interior depleted in ices, enriched ice shells - pristine ratio preserved & Pristine ratio preserved, homogeneous distribution & Heterogeneous mixture of material from depleted and enriched areas & No ices \\
            \midrule
            \textbf{Binarity, flattening and rotational orientation (Section \ref{sec:Observables:Binarity})} & Many binaries, flattened shape and aligned along the principal axis, more prograde rotating objects & Many binaries, flattened shape and aligned along the principal axis, more prograde rotating objects & Pristine binarity is destroyed, creation of new binaries after collision is possible, random alignment and distribution of inclinations & Pristine binarity is destroyed, creation of new binaries after collision is possible, random alignment and distribution of inclinations & Pristine binarity is destroyed, creation of new binaries after collision is possible, random alignment and distribution of inclinations \\
            \bottomrule
		\end{tabularx}
\end{table}

\subsubsection{Internal Morphology and Pebble Properties}
\label{sec:Observables:InternalMorhology}

\textbf{Expectation}\\
Planetesimals formed by the gentle collapse of a pebble cloud are initially very homogeneous throughout the whole body, despite the possibility of size-sorting of the pebbles while the cloud collapses~\cite{WahlbergJansson.2017}. The~latter would generate a gradient in the size of the void spaces between the pebbles, but~not necessarily in density. In~any case, the~size of the void spaces between the pebbles is always on the order of the pebble size. As~described above, compaction due to lithostatic compression of large planetesimals would decrease the void space in the inner parts of the objects. However, for~simplicity we excluded these bodies (i.e., $>$50 $\,\mathrm{km}$ in size) due to their large sizes compared to typical comets. As~pebble size is expected to be on the order of millimetres to decimetres, density contrast on length scales much larger than decimetres is not expected in the interior. For~type A1, no alteration of the internal structure due to evolutionary processes is expected, and~the internal morphology remains pristine, with~pebbles and void spaces of pebble size that are filled with fractals (see Section \ref{sec:GravitationalCollapse}). For~type B1, the~internal morphology is almost identical to A1, with~the exception that in the deep interior, the~porosity is increased due to the evaporation of ices, and~in the ice shell(s), the~pore size is reduced due to recondensation of the ice(s). If~a catastrophic collision occurred, fractals may have been lost where the void spaces were opened. Therefore, fractals may be locally depleted for type B1 and B2, but~the internal morphology of pebbles and void spaces remains unaltered because catastrophic collisions do not destroy the pebble~structure.

From the modelling point of view, pebble sizes are determined by the bouncing barrier~\cite{Guttler.2010,Zsom.2010}. For~pebbles containing microscopic water-ice particles, Lorek et al. 2018 \cite{Lorek.2018} derived maximum pebble sizes on the order of 1 mm to 1 cm, depending on the heliocentric distance and specific PPD parameters. As~to the internal morphology of the pebbles, Lorek et al. 2018 \cite{Lorek.2018} calculated filling factors between 0.3 and 0.4. Pebble substructures and possible inhomogeneous internal morphology of the pebbles have not been studied extensively. However, for~two reasons, inhomogeneity is expected: (i) pebble-growth studies show that the size--frequency distribution of the building blocks of pebbles has a certain shape and width~\cite{Zsom.2010,Lorek.2018}; if the bouncing-collision phase of the pebbles before gravitational collapse does not erase all substructures, pebbles should still be characteristically inhomogeneous. (ii) Based upon the size--frequency distribution of dust ejected into the cometary coma, Fulle et al. 2020 \cite{Fulle.2020b} and Ciarniello et al. \cite{Ciarniello.2021,Ciarniello.2022} derived a pebble-internal morphology model that is capable of explaining a variety of comet observation~facts. 

With a super-catastrophic collision (type A/B3), pebbles and fractals are destroyed, and~the resulting rubble piles are highly compacted, and~thus relatively rigid. Within~this rubble, large void spaces can remain. The~size of the rubble may vary and is not further constrained in this~work.\\

\textbf{Comparison to Observations}\\
The internal morphology of comets is only measurable by instruments with the ability to scan the interior of the objects. The~Rosetta Mission was equipped with instrument packages capable of performing these investigations, namely Comet Nucleus Sounding Experiment by Radiowave Transmission (CONSERT \cite{Kofman.2007}) and Rosetta Radio Science Investigation (RSI \cite{Patzold.2007}) instruments. Both devices performed radar observations of the interior of the~nucleus.

Their measurements showed that the observed volume of comet 67P is highly homogeneous on length-scales larger than $3.3 \,\mathrm{m}$ (CONSERT \cite{Kofman.2015}). However, CONSERT could not investigate the entire body, so the results are limited to a fraction of the volume of comet 67P. CONSERT found a denser surface layer for the uppermost $<25 \,\mathrm{m}$ \cite{Kofman.2020}. Such variation may be explained by the redistribution of volatiles due to radiogenic heating (see Section \ref{sec:Radiogenic Heating}). RSI measurements showed an offset of the centre of gravity, which indicates density difference of the two lobes. The~big lobe seemed to be slightly denser than the small lobe~\cite{Jorda.2016}.

Direct observations of the pebbles have not been unambiguously performed. Imaging of the surface of comet 67P with the highest spatial resolution was performed by the CIVA instrument onboard the Philae lander~\cite{Poulet.2016} and showed substructures with sizes of a few millimetres to one centimetre, which might be identified as pebbles. Other, indirect methods have been applied by Blum et al. 2017 \cite{Blum.2017} to derive the pebble size and result in values for the pebble radii between 3 mm and 6 mm. Cometary trails have been investigated by Lisse et al. 1998  \cite{Lisse.1998}, who found that a high abundance of macroscopic particles with \mbox{$\beta$ values} of $\beta < 10^{-3}$, corresponding to dust-particle sizes of $\gtrsim$1 $\, \mathrm{mm}$, are present in the trails. Moreover, cometary meteor showers exhibit particle sizes up to $\sim$$1 \, \mathrm{cm}$ \cite{TrigoRodriguez.2022}. As~stated above, direct observational evidence of pebble substructures is not~available.

\subsubsection{Bulk Porosity}
\label{sec:Observables:TotalPorosity}

\textbf{Expectation}\\
The bulk porosity of planetesimals is mainly influenced by formation processes and can then be altered due to evolutionary processes. As~a simple rule-of-thumb, the~packing density of a planetesimal consisting of two hierarchy levels (dust/ice particles forming pebbles; pebbles forming planetesimals) is $\Phi_\mathrm{bulk}=\Phi_\mathrm{d} \, \Phi_\mathrm{p} \approx 0.24$, with~$\Phi_\mathrm{d} \approx 0.4$ and $\Phi_\mathrm{p} \approx 0.6$ being the packing densities of the dust/ice particles inside a pebble and of the pebble packing inside the planetesimal, respectively~\cite{Skorov.2012}. For~pebbles with substructures, the~internal packing might be $\Phi_\mathrm{d} < 0.4$, as~this can be regarded as another hierarchy level~\cite{Guttler.2019}. Thus, the~bulk porosity adds up to $\sim$$76\,\%$. For~larger planetesimals, the~porosity is expected to decrease with increasing depth due to lithostatic compression of the pebble structure~\cite{Malamud.2022}. However, as~described in Section~\ref{sec:Radiogenic Heating}, radiogenic heating can affect even rather small bodies when they form early enough (type B1). The~outward diffusion of hyper- and super-volatiles goes along with a change in porosity so that the porosity increases above the primordial value close to the centre and decreases at locations where the volatiles condense. However, when no volatiles are lost into space, the~bulk porosity does not change and remains pristine. For~larger objects, a~change of porosity can also be induced due to the melting of water ice. When water ice melts, the~pebbles collapse and immerse into the water so the initial porosity is lost, and~consequently, the~body shrinks. However, as~this only happens close to the centre, the~outer layers maintain their pebbles and their porous structure. For~type B2 bodies, the~pebbles from the various regions possess different dust-to-ice ratios, which implies that the observed density across the surface and for different depths changes. For~the compacted structures of type A/B3, very little porosity remains inside the rubble, but~macroporosity may exist within it. Thus, the~bulk porosity of the entire body of a type A/B3 should be lower than the pristine~value.  \\

\textbf{Comparison to Observations}\\
The porosity of a whole body can be derived from its volume, mass and material density. Material composition and porosity can be estimated by measurement of the electric permittivity~\cite{Herique.2019}, but~this procedure is model-dependent and relies on calibration experiments~\cite{Brouet.2016}. Measurements of total porosity with high resolution were only performed for comet 67P, resulting in values for the bulk packing density of $\Phi_\mathrm{bulk}=0.15$ to $\Phi_\mathrm{bulk}=0.37$. The~RSI instrument measured values between $\Phi_\mathrm{bulk}=0.25$ and \mbox{$\Phi_\mathrm{bulk}=0.30$ \cite{Patzold.2016}}. An~estimate of $\Phi_\mathrm{bulk}=0.21$ to $\Phi_\mathrm{bulk}=0.37$ was retrieved by directly measuring the pebble density by Grain Impact Analyser and Dust Accumulator (GIADA \cite{Colangeli.2007,Fulle.2016b}). Moreover, permittivity measurements done by CONSERT provided a range of $\Phi_\mathrm{bulk}=0.15$ to $\Phi_\mathrm{bulk}=0.25$ \cite{Kofman.2015}. The~lower value of $\Phi_\mathrm{bulk}=0.15$ leaves room for a third hierarchy level, i.e.,~for pebbles with substructures, whereas the upper value requires the pebbles to be~homogeneous. 

Due to the overall high porosity of comet 67P, evolutionary processes that considerably reduce the porosity, such as, e.g.,~super-catastrophic collisions, can be~excluded.

\subsubsection{Dust-to-ice Mass Ratio}
\label{sec:Observables:DIR}

\textbf{Expectation}\\
The dust-to-ice mass ratio (DIMR) of a planetesimal is given by the material distribution inside the pebble cloud that undergoes the gravitational collapse. The~DIMR depends on the location of formation as well as on the preceding mixing processes of the materials during pebble growth. We refer to the DIMR at cloud collapse as pristine, but~it should be noted that this DIMR is not necessarily equal to the DIMR of the solar nebula or of the bulk protoplanetary disc outside the various snowlines. During~the thermal and collisional evolution of the planetesimals, the~DIMR can be~altered.

For comet types A1 and A2, the~pristine DIMR is globally and locally preserved. Throughout the entire volume, the~distribution of ices is homogeneous down to the pebble scale. For~type B1, the~bulk abundances of all ices are preserved, but~the local DIMR is changed. The~interior of the planetesimal is depleted in ices (higher DIMR), and~the ice layers are enriched in ices (lower DIMR). The~near-surface layer remains mostly pristine in the DIMR, despite locations where material was ablated by collisions. As~type B2 is the breakup product of a differentiated body, the~distribution of ices is inhomogeneous and depends on the depth of the place inside the body where the rubble was originally stored. The~manner in which such a body re-accumulates the collisional fragments determines the distribution of the ices~\cite{Malamud.2022}. Whether the DIMR of the rubble is changed compared to the pristine value depends on the details of the preceding catastrophic collision event. For~types B1 and B2, super-volatiles could be lost entirely if the corresponding threshold temperature was reached at the moment of the collision. Type A/B3 does not contain any ices due to the hyper-velocity nature of the collision~event.\\

\begin{figure}
\centering
\includegraphics[width=14 cm]{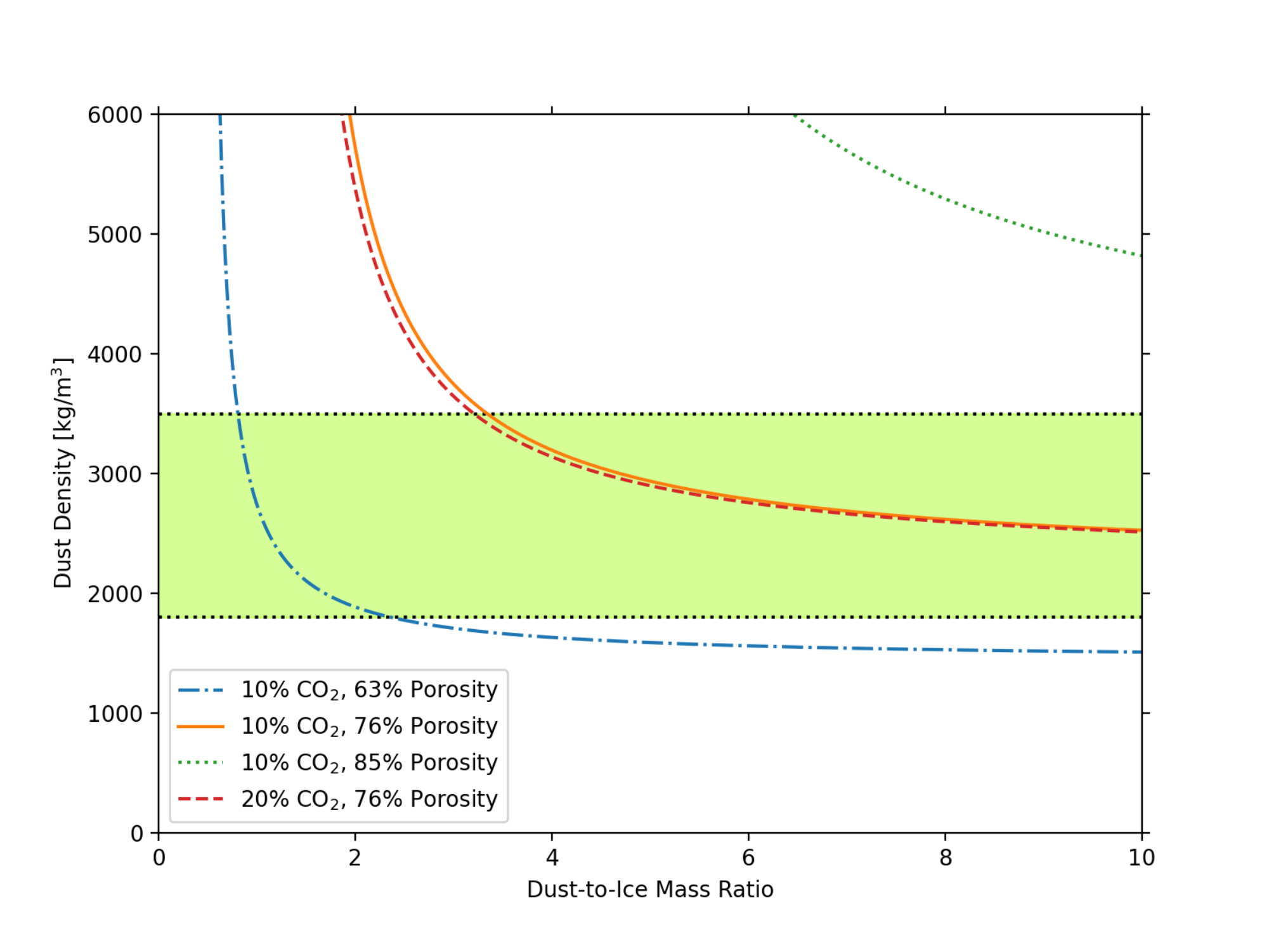}
\caption{This plot shows the resulting dust density for varying dust-to-ice mass ratios and porosities. The green-coloured area indicates reasonable dust densities. The given CO$_2$ percentage corresponds to the CO$_2$ content of the total ice mass. \label{fig:dust_density}}
\end{figure}

\textbf{Comparison to Observations}\\
The dust-to-ice mass ratio $F$ can also be addressed by porosity and density estimations, as~shown by Lorek et al. 2016 \cite{Lorek.2016} and Patzold et al. 2019 \cite{Patzold.2019}. For~cometary refractory material, the~material density is not known and can only be constrained within a range of reasonable values. Measurements by Rosetta found an organic-to-mineral mass ratio of roughly 1:1~\cite{Bardyn.2017}. Typical densities of organic matter (1000--2000  
 $\,\mathrm{kg/m^3}$), combined with typical densities of minerals (2600--5000 $\,\mathrm{kg/m^3}$), result in an expected density range of \mbox{1800--3500 $\,\mathrm{kg/m^3}$} for the refractory material of a comet. GIADA measured comparable densities with $1925^{+2030}_{-560} \,\mathrm{kg/m^3}$ \cite{Fulle.2017b}. Some particles showed a bulk density >4000 $\,\mathrm{kg/m^3}$. For~water and CO$_2$ ice, the~densities are  $\rho_{\mathrm{H_2O}} = 934 \,\mathrm{kg/m^3}$ and $\rho_{\mathrm{CO_2}} = 1600 \,\mathrm{kg/m^3}$, respectively. Porosity estimations for comet 67P are shown in Section~\ref{sec:Observables:TotalPorosity}. With~this, one can calculate the dust density $\rho_d $ with:
\begin{equation}
    \rho_d = \frac{f_d \, \rho_{\mathrm{n}}}{ 1 - f_{\mathrm{H_2O}} \frac{\rho_{\mathrm{n}}}{\rho_{\mathrm{H_2O}}} - f_{\mathrm{CO_2}} \frac{\rho_{\mathrm{n}}}{\rho_{\mathrm{CO_2}}} - \Phi }
\end{equation}
with $f_d + f_{\mathrm{H_2O}} + f_{\mathrm{CO_2}} = 1$ the mass fractions of dust, water and CO$_2$, respectively, \mbox{$\rho_{\mathrm{n}} = 532 \,\mathrm{kg/m^3}$} the nucleus density~\cite{Jorda.2016} and bulk porosity $\Phi$.
The results for varying DIMR are shown in Figure~\ref{fig:dust_density}. From~this plot, one can constrain the global porosity and the ice content. Porosities larger than $\sim$$81 \%$ cannot be explained with reasonable dust densities and should therefore be excluded. For~a porosity of $76\%$ (see Section \ref{sec:Observables:TotalPorosity}), the~DIMR has a range of 3 to 4 depending on the assumed CO$_2$ fraction. For~the smallest porosity of $ 63\%$ derived from Rosetta instruments, only a small range of DIMR values between $0.8$ and $2.5$ is reasonable. 
\par
These values are in agreement with the review written by Chourkoun et al. 2020 \cite{Choukroun.2020} about the refractory-to-ice(s) mass ratio of comet 67P. In~this work, the~authors showed, based on different Rosetta observations, that the DIMR is most probably higher than 3 (see Figure~5b in \cite{Choukroun.2020}).  
An~upper limit of the DIMR, which is in agreement with most of the Rosetta measurements, can be between 6 to 8. Only two papers argue for much higher maximum values. The~works claiming higher DIMR values are based on coma dust grain density estimation around perihelion~\cite{Fulle.2017b} and on the analysis of the fallback of volatile-bearing chunks in the northern hemisphere~\cite{Fulle.2019b}.

\subsubsection{Binarity, Flattening and Rotational Orientation}
\label{sec:Observables:Binarity}
\textbf{Expectation}\\
Numerical simulations have shown that gravitationally unstable pebble clouds can collapse into binary planetesimals. These objects rotate in a prograde manner in $80\%$ of the simulations~\cite{Nesvorny.2019}. This finding is almost not affected by different parameters influencing streaming instability. Such a formation of binaries results in very similar compositions of both bodies because they form from the same reservoir of material. The~radii of both bodies would also be very similar~\cite{Fraser.2017}. The~spinning of the planetesimals induced by the collapse results in a flattened shape~\cite{McKinnon.2020}. Additionally, the~ 
parts of such a binary align over time because this reduces the energy of the system. The~alignment can be followed by a reduction of the distance between both bodies until they contact due to, e.g.,~Kozai--Lidov cycling, the~YORP and BYORP effects, tides, collisions or gas drag~\cite{McKinnon.2020}. However, the~collapse of a pebble cloud can lead to a wide variety of binary distances, including  contacting binaries~\cite{Nesvorny.2010}. The~survival of binaries in collisions was addressed by Nesvorny et al. 2019 \cite{Nesvorny.2019b}, and~they found that more tightly bound binaries have a higher survival probability than wider ones. Regarding radiogenic heating, due to the small contact area compared to the total area, the~influence is negligible and should be equal for each body individually. It should be mentioned that the binarity of low-density objects also has important consequences for the escape of dust particles when dust activity is present (see, e.g.,~\citep{Wang.2014,Jiang.2017}).

\textbf{Comparison to Observations}\\
Many of the observed cometary nuclei show a binary shape. The~most detailed observed one, comet 67P, exhibits this in great detail. However, the~formation of binary shapes is still discussed and researched. As~shown before, the~formation of binaries is an intrinsic characteristic of formation by a collapsing pebble cloud. Nesvorny et al. 2019 \cite{Nesvorny.2019} showed that for binaries that have formed by capturing another object, the~distribution of inclinations would favour retrograde over prograde rotations. Comparison to the known binaries in the Kuiper belt shows that the distribution of binary inclinations can be best explained by direct formation of binaries from a single source~\cite{Nesvorny.2019}. Arrokoth shows alignment of the two lobes, which is in agreement with the predicted scenario~\cite{McKinnon.2020}. However, for~comet 67P, the~observation of layered surface structures surrounding each lobe~\cite{Massironi.2015,Penasa.2017,Ruzicka.2019} suggests that the lobes formed individually and contacted~later.

\subsection{Derived Physical Properties}
\label{sec:Discussion:Derivative_Physical_Properties}

\begin{table}[]
\centering
\caption{This table presents an overview of the different proposed types of objects with their expected derived physical properties, which are discussed in detail in the following sub-sections.\label{tab:overview_body_types_derivative_properties}}
		\newcolumntype{C}{>{\centering\arraybackslash}X}
		\begin{tabularx}{18cm}{CCCCCC}
			\toprule
				& \textbf{Type A1}	& \textbf{Type B1}     & \textbf{Type A2} & \textbf{Type B2} & \textbf{Type A/B3}\\
			 &   \includegraphics[width=1cm]{subcatastrophic_undifferentiated_updated.pdf} &   \includegraphics[width=1cm]{subcatastrophic_differentiated_updated.pdf} &  \includegraphics[width=1cm]{icy_rubble_Pebble_pile_updated.pdf} &  \includegraphics[width=1cm]{icy_rubble_Pebble_pile_diff_updated.pdf} &  \includegraphics[width=1cm]{non_icy_rubble_pile_blue.pdf} \\
			 \textbf{Derived physical properties} & \multicolumn{2}{c}{Icy pebble piles} & \multicolumn{2}{c}{Icy rubble/pebble piles} & Non-icy rubble piles \\
            \midrule
            \textbf{Cometary activity (Section \ref{sec:Observables:Activity})} \textit{Instruments: Camera, dust collector, gas detector} & Possible everywhere at the surface, activity will not change when eroded & Possible everywhere at the surface, but activity changes when eroded & Possible everywhere at the surface, but activity changes when eroded & Possible at specific locations, erosion can change activity patterns & Not possible \\
            \midrule
            \textbf{Thermal conductivity (Section \ref{sec:Observables:ThermalConductivity})} \textit{Instruments: Thermal IR, microwave, in situ temperature sensors} & Low network and radiative conductivity at night, radiative part strongly increases at day due to strong temperature dependence & High radiative thermal conductivity in the interior, the shells have a moderately increased network conductivity, and the mantle material behaves like Type A1 & Low network and radiative conductivity at night, radiative part strongly increases at day due to strong temperature dependence & Different locations are either dominated by the low network, or the varying radiative thermal conductivity & Relatively high network conductivity inside the rubbles, the large voids in between are dominated by radiation\\
            \midrule
            \textbf{Tensile strengths (Section \ref{sec:Observables:TensileStrength})} \textit{Instruments: e.g. camera (cliff collapse)} & Ultra low intra-pebble tensile strength and low inter-pebble tensile strength & Ultra low intra-pebble tensile strength and low inter-pebble tensile strength, both enhanced in ice-enriched areas and decreased at in ice-depleted areas & Ultra low intra-pebble tensile strength and low inter-pebble tensile strength & Ultra low intra-pebble tensile strength and low inter-pebble tensile strength, both enhanced in ice-enriched areas and decreased at in ice-depleted areas &  High intra-rubble tensile strength, but almost no tensile strength between the rubbles\\
            \midrule
            \textbf{Compressive strengths (Section \ref{sec:Observables:CompressiveStrength})} \textit{Instruments: Measurement of impact decelleration} & Low pressures required to compress the pebble material & Low pressures required to compress the material in the mantle, the pressure required for compression is increased in the ice shells and decreased in the depleted interior & Low pressures required to compress the pebble material & Varying compression response of the material, depending on the ice fraction & Rubbles require more compression pressure to be compacted to the same volume filling factor as the pebbles \\
            \midrule
            \textbf{Gas permeability (Section \ref{sec:Observables:GasPermeability})} \textit{Instruments: Gas detector} & High permeability inside the large voids between the pebbles and low permeability inside the pebbles & Gas permeability is decreased in layers of enhanced ice content (ice shells), gas permeability is decreased where lice is less abundant & High permeability inside the large voids between the pebbles and low permeability inside the pebbles & Varying gas permeability at different locations at the surface and in the interior & Very low gas permeability  \\
            \midrule
            \textbf{Permittivity (Section \ref{sec:Observables:Permittivity})} \textit{Instruments: Deep penetrating radar, DC geoelectric} & Homogeneous electrical permittivity on the surface and in the interior & Varying electrical permittivity for different parts of the nucleus (mantle, shells, interior) & Homogeneous electrical permittivity on the surface an in the interior & Inhomogeneous electrical permeability on the rubble scale, but homogeneous inside the rubbles & Homogeneous electrical permittivity inside the rubbles, but variations because of the large voids \\
            \bottomrule
		\end{tabularx}
\end{table}

\subsubsection{Cometary Activity}
\label{sec:Observables:Activity}
\textbf{Expectation}\\
If we define cometary dust and gas activity as a state in which insolation produces subsurface pressures large enough to overcome the cohesion of the dust, then types A1, A2, B1 and B2 can in principle become active due to the presence of volatile ices close to the surface. However, these objects show significant differences, which are noteworthy and could become decisive to answer the question of how comets formed and evolved. The~distribution of volatiles over the entire surface of the bodies is different. Due to the homogeneity of types A1 and A2, these objects should become active at each surface location given sufficient insolation. When the surface of a type-B1 object gets eroded locally, the~activity can be different at this location because an ice shell or the depleted interior could be excavated. We could observe this effect by local variations of the outgassing rates of different volatile species. For~type B2, the~different ice-containing rubble is distributed inhomogeneously over the whole body. Hence, also for type B2 we expect locally varying outgassing rates. Type A/B3 cannot show cometary activity driven by ice sublimation as it no longer contains any~ices.  \\

\textbf{Comparison to Observations}\\
The first-ever observed cometary nucleus was the nucleus of comet 1P/Halley. Halley was expected to show global activity, but~when the Giotto spacecraft arrived, only $\sim$$1/3$ of the surface was contributing to global activity~\cite{Keller.1987}. A~similar observation was made for comet 19P/Borrelly --- one distinct jet originated from the cometary nucleus, which actually consisted of three smaller outbursts~\cite{Soderblom.2004}. In~contrast to other comets, comet 81P/Wild 2 possessed a diverse and complex variety of surface structures~\cite{Tsou.2004}. Longer-exposure camera images revealed the presence of many jets around the entire nucleus. More than twenty highly collimated jets were observed, which indicated a heterogeneous surface. In~addition, analysis of captured particles indicated that comet Wild 2 may have experienced aqueous alteration in the past~\cite{Berger.2011}. Comet 9P/Tempel 1 showed rather variable dust activity, with~frequent natural outbursts all over the surface~\cite{AHearn.2005}. In~comparison to the above-mentioned comets, 103P/Hartley 2 had very distinct activity patterns. During~the flyby of the EPOXI spacecraft~\cite{AHearn.2011b}, water vapour emission was detected at the waist of nucleus. Localised CO$_2$-active areas also showed ejection of solid water ice chunks of up to cm size. In~addition, solid ice particles were also observed in an area where only water was active.
\par
Comet 67P is by far the best-studied comet, and~this is why we are aware of several local activity hotspots (such as sunset jets; see, e.g.,~\cite{Shi.2016}), although~the nucleus seems to be globally active~\cite{Lauter.2018}. The~activity pattern changed with time because of seasonal variations of solar illumination and because of material that fell back onto the nucleus~\cite{Fulle.2019b}. During~perihelion, the~southern hemisphere showed an increase of CO$_2$ activity that led to the ejection of larger ice-containing chunks (up to several cm in size~\cite{Ott.2017}) and to a bluing of the surface because more water-ice-containing pebbles were excavated by CO$_2$ erosion (for details see \citep{Ciarniello.2022}). These chunks either escaped into space or fell back into the northern hemisphere of the nucleus. Although~we understand the principle of how gas activity leads to the ejection of pebbles or chunks from the surface, it is still not known how to explain the particle-size distribution in the coma~\cite{Blum.2017}. The~smallest (sub-)micrometre-sized particles cannot be lifted directly from the surface because of their strong adhesion forces. This means that the chunks or pebbles must disintegrate in the coma after lift-off to produce the observed fraction of small particles in the coma. Fulle et al. 2020 \cite{Fulle.2020} proposed that pebbles possess a substructure called agglomerates, typically ranging from micrometre to millimetre in size. Probably charging effects, fast rotation of the pebbles or extreme illumination conditions lead to break up of the pebbles into unbound agglomerates with a size distribution similar to the observed~one.

\subsubsection{Thermal Conductivity}
\label{sec:Observables:ThermalConductivity}
\textbf{Expectation}\\
Heat can be transported by conduction, radiation, gas diffusion and convection in the case of fluid material. The~latter mechanism can be neglected in contemporary comets because they have certainly cooled down, even if they had at some point reached the critical temperature for water-ice melting~\cite{Malamud.2022}. Further, heat transport by gas diffusion can be neglected in most cases due to the low gas pressures involved. For~the conduction and radiation of thermal energy, material and structural properties determine the efficiency of heat transport. In~a granular medium, conduction is less effective than in solid material due to the reduced contact areas between neighbouring particles, which is described by the so-called Hertz factor. For~pebbles, heat conduction is reduced further because of the granularity inside the pebbles. Therefore, and~due to the large void spaces between the pebbles, thermal radiation contributes to the energy transport and can even dominate for large pebbles and high temperatures, as~shown in Bischoff et al. 2021 \cite{Bischoff.2021} (their Figure~1). 
 Only for low temperatures (<100 $\,\mathrm{K}$ for pebbles with $5 \,\mathrm{mm}$ radius) does heat conduction dominate, but~then the thermal conductivity is extremely small (<$10^{-3} \, \mathrm{W/(K \, m)}$).

For the five types of bodies proposed, this means the following: Types A1 and A2 with their pebbles show low network conduction, and~radiation dominates for higher temperatures. For~type B1, the~shell structure influences the conductivity such that the interior with reduced ice content shows decreased network conductivity compared to the pristine material due to the higher porosity and the low internal temperatures. As~the ice shells are enriched with material, their network conductivity is locally increased. For~the outer pristine material, the~network conduction is the same as for types A1 or A2. For~type B2, which possess regions of different dust-to-ice ratios, thermal conductivity will vary among the different locations. Ice-depleted areas, which possess higher porosities, are subject to strong day--night thermal conductivity variations. Areas of higher ice content generally possess higher thermal conductivity because of the network part, but~temperature changes do not have the same importance as for the ice-depleted parts. In~case of type A/B3, the~rubble is expected to be compacted, and~therefore, network conduction is the relevant transport~process.\\

\textbf{Comparison to Observations}\\
Instead of the thermal conductivity $\lambda$, often the thermal inertia $\Gamma= \sqrt{\lambda \rho c}$ is measured for small solar system objects, with~$\rho$ and $c$ being the mass density and heat capacity of the material, respectively. However, these properties are often not known with high accuracy, and~additionally, the~temperature dependency of all these values is often not taken into account. As~described above, the~temperature dependency can be crucial for highly porous pebble-structured materials. Thus, measured values of the thermal conductivity or thermal inertia should be considered with caution and in the corresponding temperature range. For~comet 67P, several instruments onboard Rosetta and Philae were able to measure temperatures. On~the orbiter, Visible InfraRed and Thermal Imaging Spectrometer \mbox{(VIRTIS) \cite{Coradini.2007}} measured the surface temperature at depths of tens of micrometres. Self-heating and shadowing effects were observed by VIRTIS~\cite{Tosi.2019}. However, VIRTIS had a lower limit for temperature detection of $\sim$$156 \,\mathrm{K}$ due to instrument noise. The~Microwave Instrument for the Rosetta Orbiter (MIRO) \cite{Gulkis.2007} used wavelengths of $0.5 \,\mathrm{mm}$ and $1.6 \,\mathrm{mm}$, resulting in penetration depths on the order of a few millimetres to centimetres~\cite{Blum.2017}. The~MIRO temperature data are difficult to interpret because the measurement depth is comparable to the diurnal skin depth~\cite{Gulkis.2015,Schloerb.2015}, and~therefore a wide temperature range contributes to the measured brightness temperature. Marshall et al. 2018 \cite{Marshall.2018} found a best-fitting thermal inertia of $80 \,\mathrm{J/(K\, m^2\, s^{0.5})}$ under the assumption of a complex roughness distribution when comparing MIRO and VIRTIS data. Onboard the lander Philae, Multipurpose Sensors for Surface and Subsurface Science (MUPUS) \cite{Spohn.2007} was capable of probing the surface with the MUPUS PEN (thermal probe) and the MUPUS TM (16 resistance temperature detectors and an infrared radiometer). They found a best-fitting thermal inertia of \mbox{$85 \,\mathrm{J/(K\, m^2\, s^{0.5})}$ \cite{Spohn.2015}}. Assuming a heat capacity of $560 \,\mathrm{J/(kg\, K)} $ (as used in Bischoff et al. 2021 \cite{Bischoff.2021}) and a density of $532 \,\mathrm{kg/m^3} $ (for the bulk nucleus,~\cite{Jorda.2016}), these thermal inertia values translate into thermal conductivities of $\sim$$0.02 \,\mathrm{W/(K \,m)}$. As~described before, the~uncertainty in heat capacity and density of the measurement spots is forwarded into uncertainties of the thermal conductivity. Thus, such a value alone cannot be used to conclude the kind of structure of the cometary surface. The~method proposed by Bischoff et al. 2021 \cite{Bischoff.2021}, which allows the distinction between micro- and macro-porous subsurface structures by relating the sunrise temperatures to the insolation at noontime, could not be used with measured data of comets due to the lower limit of available temperature data. However, in~future missions to comets, such as ESA’s Comet Interceptor~\cite{Snodgrass.2019}, the~application of this method should be kept in mind when measuring surface~temperatures.

\subsubsection{Tensile Strength}
\label{sec:Observables:TensileStrength}
\textbf{Expectation}\\
The cohesion of granular matter is in general influenced by the material composition, its structure and the observed length scale. The~tensile stress a body can withstand before it breaks is called the tensile strength. For~non-porous solid material, the~tensile strength is highest due to the direct contact of the atoms and is typically found in the MPa range. Tensile strength is reduced by any kind of porosity, granularity or roughness because of the reduction of contact area among the atoms. For~objects composed of micrometre-sized grains, i.e.,~for the inside of pebbles, the~tensile strength is on the order of kPa~\cite{Gundlach.2018b,Bischoff.2020}. For~pebble-piles, the~tensile strength is further reduced to a few Pascals or even less, depending on pebble size~\cite{Skorov.2012,Blum.2014,Brisset.2016}. For~planetesimals formed from pebbles, such a low tensile strength is expected on the metre-scale or above. Compaction due to collisions or redistribution of volatiles leads to higher tensile strength values. The~relation between volume-filling factor and tensile strength has been measured for several materials, such as silica, graphite and organics, and~depends strongly on the compaction~\cite{Gundlach.2018b,SanSebastian.2020,Kimura.2020b,Bischoff.2020}. 
Applying this knowledge to the five types of bodies discussed at the bottom of Figure~\ref{fig:overview} leads to the conclusion that types A1 and A2 possess low tensile strength on the order of 1 Pa down to the length scale of the pebble size. As~shown in Section~\ref{sec:CollisionalEvolution}, sub-catastrophic and catastrophic collisions can occur without causing significant damage to the pebble structure. Hence, the~tensile strength is not affected by collisional evolution in these two cases. However, differentiation causes significant changes of the tensile strength. Locations of enhanced ice content will possess higher tensile strength values, whereas depleted volatile material consequently leads to a decrease in tensile strength. This is why the redistribution of material in the rubble in type B2 leads to variation of the tensile strength over the surface and volume of these bodies. The~material of type A/B3 bodies is highly compacted, and~this is why the rubble should have a high intra-rubble tensile strength. However, the~tensile strength between the rubble particles is almost zero, which means that the rubble is only bound by~gravity. 

\textbf{Comparison to Observations}\\
As described above, the~tensile strength of a body varies for different length scales. For~large scales of several metres to 100 m, the~tensile strength of comet 67P was estimated by the observation of cliff collapses~\cite{Attree.2018} and found to be on the order of 1 Pa, as~expected for a pebble-pile body. For~length scales of a kilometre or 
above, the~tensile strength is difficult to measure because of the increased importance of gravitational binding. The~central hydrostatic pressure of comet 67P, for~example, is on the order of 10 Pa, and~thus larger than the expected tensile strength of a loose pebble pile. However, for~bodies with radii of several kilometres or above, a~memory effect can lead to increased tensile strength as shown by Blum et al. 2014 \cite{Blum.2014}. These results may indicate why the lower limits of the tensile strengths for a number of comets are slightly above the canonical 1 Pa value~\cite{Blum.2006}. Much smaller length scales can be probed through the breakup of cometary meteors in Earth’s atmosphere. These typically millimetre-sized bodies possess tensile strengths on the order of 1--10 kPa~\cite{Blum.2014} and may thus represent intact pebbles or fragments thereof.

\subsubsection{Compressive Strength}
\label{sec:Observables:CompressiveStrength}
\begin{figure}
\centering
\includegraphics[width=13.79cm]{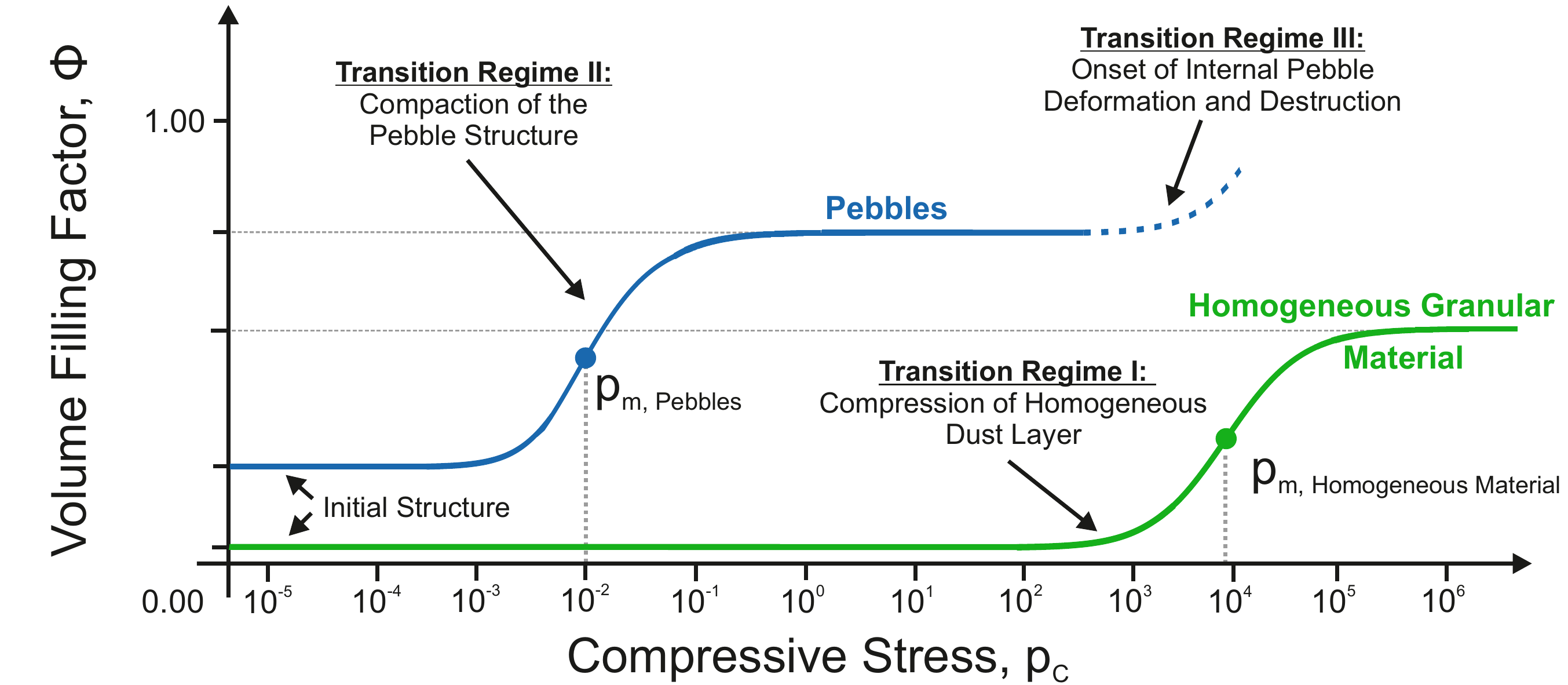}
\caption{Response function of 
 a granular material to compressive stress. The~green curve shows the reaction of a homogeneous, i.e.,~non-hierarchical, granular material, whereas the blue curve represents the volume-filling factor of a pebble-pile structure, i.e.,~a hierarchical material. Both functions can be described by S-shaped curves as shown by \cite{Schrapler.2015}. 
 The~homogeneous material can only be compressed in a single stage (transition regime I), while for the hierarchical material, first the pebble packing will be compressed (transition regime II), then at higher pressures the pebbles will be deformed and destroyed (transition regime III; dashed curve). \label{fig:compression}}
\end{figure}

\textbf{Expectation}\\
Compressive strength describes the ability of a material to withstand pressures tending to reduce size. For~granular materials, this material property depends on the volume-filling factor of the material as well as on the grain properties, such as shape and size distribution. This is why the compressive strength of a granular medium cannot be seen as a single value, but~should rather be considered a volume-filling-factor-dependent value that changes with respect to the applied compressive stress (see Figure~\ref{fig:compression}). The~reason is that fluffier materials can be compacted much easier in comparison to more compact structures. The~functional behaviour of the compression curve of granular materials was derived using theoretical considerations and laboratory experiments~\cite{Schrapler.2015}. The~result is that the response of the volume-filling factor to compressive stress can be described by an S-type function (see Figure~\ref{fig:compression}), with~a transition regime that is characterised by the so-called turnover point, $p_m$. Homogeneous granular materials are compressed in the transition regime I. Larger pressures cannot lead to any further compaction because the grains are already positioned in the densest possible packing. For~pebble piles, however, the~compression function can be separated into two different regimes, namely transition regime II, in~which the pebbles are brought into a denser packing without destruction~\cite{Malamud.2022}; and transition regime III, in~which the pebbles cannot withstand the applied pressure and are, hence, destroyed~\cite{ORourke.2020}.

For type A1 and A2 bodies, low-stress compressive strength should be low and comparable to the tensile strength. For~types B1 and B2, the~general behaviour should be similar, but~due to the enhanced volume-filling factor at the ice shell, the~compressive strength should be increased, and~inhomogeneously distributed in type B2. The~compact rubble in bodies of type A/B3 possess a high compressive strength on length scales smaller than the rubble~size. \\

\textbf{Comparison to Observations}\\
The only available estimations of the compressive strength come from the landing attempts of Philae on the nucleus of comet 67P. Heinisch et al. 2019 \cite{Heinisch.2019} determined an upper limit of 800 Pa, while O'Rourke et al. 2020 \cite{ORourke.2020} derived an upper limit of only 12 Pa for the compressive strength. Based on the compression model for granular matter under reduced gravity conditions by Schräpler et al. 2015 \cite{Schrapler.2015}, these low values can be interpreted as being due to loosely packed pebble-sized bodies on the surface of comet 67P~\cite{ORourke.2020}.

\subsubsection{Gas Permeability}
\label{sec:Observables:GasPermeability}
\textbf{Expectation}\\
The gas permeability of a small solar system body is determined by its porosity and the characteristics of the pore spaces, such as their size distribution and percolation properties. Tortuosity describes the actual length of the path the gas takes divided by the closest distance between the start and end points. Hence, high tortuosity hinders the gas flow, whereas low values of tortuosity provide favourable gas-flow conditions. For~planetesimals, the~gas permeability therefore depends on the porosity distribution inside the body. For~larger pebbles, the~gas permeability increases. Therefore, bodies of types A1 and A2 possess relatively high gas permeability. However, if~the redistribution of volatiles due to radiogenic heating results in filling the voids in the ice shell (types B1 and B2), the~gas permeability can be greatly reduced, and~the flow could even be stopped if the pores are closed completely. In~the deep interior, the~permeability could be increased where icy pebbles are extinct, but~this depends on whether there are pebbles consisting only of ices, leaving large voids behind, or~pebbles that are mixtures of ice and dust, with~the voids created by ice evaporation being small. For~type B2 bodies, the~permeability is inhomogeneous and follows the distribution of rubble from the interior, the~ice shell or the pristine mantle. As~type A/B3 bodies do not contain any ices which could sublimate, gas permeability is not of interest. In~general, the~gas permeability directly influences the possible pressure build-up inside a body with evaporating volatiles, which can lead to ejection of material if the cohesion is~overcome.

\textbf{Comparison to Observations}\\
Measurements of the gas permeability of the cometary subsurface material are not directly possible. However, using thermophysical models for the interpretation of the gas production rates of comets may shed light into the vapour transport inside the cometary nucleus. This was done by, e.g.,~Gundlach et al. 2020 \cite{Gundlach.2020}, who found that a pebble pile can explain the observations of the dust and gas emission of comet 67P around perihelion. Skorov et al. 2021 \cite{Skorov.2021} investigated how variations of porosity and inhomogeneities of the surface material affect the outgassing behaviour. They found high influence due to porosity changes, but~only minor contributions due to cavities and cracks. Skorov et al. 2022 \cite{Skorov.2022} showed that heterogeneous microstructures of dust layers can be neglected, so that effectively a homogeneous dust layer is a valid assumption for modelling outgassing. Further modelling of the transport characteristics was performed by Reshetnyk et al. 2021 \cite{Reshetnyk.2021}, which is useful for the application to comet~observations.

\subsubsection{Permittivity}
\label{sec:Observables:Permittivity}
\textbf{Expectation}\\
Permittivity is a measure of the electric polarisability of a material in response to an electric field and is dependent on the frequency of the electromagnetic wave, porosity, temperature and composition of the body. It can be divided into a real and an imaginary part. Brouet et al. 2014 \cite{Brouet.2014} investigated the permittivity of porous granular matter in different grain size ranges. Besides~refractory material, water ice can have a large influence due to the strong dependency of the permittivity on frequency. For~frequencies below $10^4 \,\mathrm{Hz}$, the~permittivity $\epsilon$ is roughly three orders of magnitudes higher than above $10^4 \,\mathrm{Hz}$  \cite{Artemov.2019}. This effect could be used to investigate the water-ice abundance. Due to the homogeneity of objects of types A1 and A2, the~permittivity is homogeneous throughout the whole body. The~differentiation in type B1 objects results in a radial gradient of the permittivity. As~describe above, water ice and silica can have similar permittivities in some frequency ranges. Therefore, distinction between the two is only possible in frequency regimes in which the materials possess different permittivities. For~objects of type B2, the~permittivity is distributed inhomogeneously, following the rubble material composition. As~A/B3 bodies are by definition ice-free, there is no influence due to water-ice permittivity, but~the permittivity is expected to be homogeneous inside single~rubble.     \\

\textbf{Comparison to Observations}\\
With the Comet Nucleus Sounding Experiment by Radiowave Transmission (CONSERT) \cite{Kofman.2007}, the~permittivity of comet 67P was investigated with a frequency of \linebreak\mbox{90 MHz~\cite{Herique.2016,Herique.2019,Kofman.2020}}. The~relative permittivity ranged from 1.7 to 1.95 in the subsurface ($<$25 m) and from 1.2 to 1.32 in the interior. A~denser surface layer or a difference in composition could result in this dichotomy~\cite{Kofman.2020}, which could have evolved due to activity and redistribution of volatiles. The~interior seems to be more porous. This characteristic is similar to that of a type B1 body, but~the length scales could be different, and~more detailed simulations are needed to address this point. Further, the~Surface Electric Sounding and Acoustic Monitoring Experiment (SESAME) \cite{Seidensticker.2007} instrument package, including the permittivity probe onboard Philae, found hints of a more compacted surface layer compared to the interior~\cite{Lethuillier.2016}; however, the~instrument only probed the first metres. Additionally, MIRO data can be used to address the material properties, as~performed by Bürger~et~al. (subm.) \cite{Burger.2022} for optical constants in the millimetre-wavelength range, which are linked to the permittivity. They found that a pebble-structured surface model can fit the measured MIRO data. In~general, the~Rosetta results of permittivity suggest high homogeneity, similar to bodies of type A1 and~A2.

\section{Conclusions and Open Questions}
\label{sec:Conclusion}
In this review, we presented our current understanding of planetesimal formation and evolution in the context of comets being possibly the most primitive survivors of these eras. We illustrated the most crucial processes and pathways from protoplanetary dust to contemporary bodies of the solar system in Figure~\ref{fig:overview}. As~ices are an important ingredient of comets, comet formation as macroscopic bodies must start beyond the snowlines of the volatiles and super-volatiles found in comets, e.g.,~CO$_2$ and CO. Through three phases of coagulation (see Figure~\ref{fig:overview} and Section \ref{sec:Coagulation}), pebbles consisting of microscopic ice and dust grains form in the size range of millimetres to decimetres and cannot grow further due to growth barriers. These pebbles get concentrated by hydrodynamic effects, leading to streaming instability, which enhances the concentration of pebbles to the gravitational-instability limit, followed by a gentle collapse of the pebble cloud into a many-kilometre-sized object. These planetesimals follow a broad size--frequency distribution, but~to evolve into comets, their size cannot exceed 50 km due to the lithostatic pressure, which destroys the pebbles for larger planetesimal sizes. Present numerical models of the collapse stage do not have the resolution to predict frequency at the small end of the expected size distribution of planetesimals, particularly for sizes in the 1-10 km size range applicable to comets. Here, more numerical work is needed to investigate the expected properties of the smallest~planetesimals. 

The first phase of planetesimal evolution is dominated by radiogenic heating by short-lived radioactive nuclei, which can redistribute volatiles inside the planetesimals. This phase is followed by collisional evolution, whose outcome depends on the specific impact energy. In~this review, we differentiate between sub-catastrophic, catastrophic and super-catastrophic collisions. All these evolutionary processes result in five distinct types of evolved planetesimals, designated by us as types A1, A2, B1, BS and A/B3 (see the bottom of Figure~\ref{fig:overview}). Additionally, it should be kept in mind that thermal alteration by solar illumination on long time scales regarding the path from the Kuiper Belt or the scattered disc inwards can influence the abundances of volatiles~\cite{Gkotsinas.2022}.

The Rosetta Mission to comet 67P Churyumov--Gerasimenko provided a huge number of scientific results as a baseline for the study of comet formation and evolution. However, there are still many open questions, and~it should be kept in mind that these detailed measurements are only available for one comet whose representation of the class of all comets in general is unknown. Only objects of type A/B3 as non-icy rubble piles can be excluded to represent a contemporary comet. Icy pebble piles (objects of types A1 and B1) preserve most of the pristine planetesimal properties and would be most interesting for the understanding of planetesimal formation. However, the~possibilities to distinguish icy pebble piles from icy rubble/pebble piles (types A2 and B2) are restricted. As~shown in Tables~\ref{tab:overview_body_types_intrinsic_properties} and \ref{tab:overview_body_types_derivative_properties}, the~only distinction between types A1 and A2 seems to be the remaining amount of fractal dust aggregates and the binarity.  Additionally, the~expected differences between the various types need to be modelled numerically, as~here we have only made a relatively generic description of their physical properties. Hence, simulations of collisional outcomes of colliding pebble piles are urgently required, including features such as the survival of fractals in the void spaces between the pebbles and differentiation of the bodies following heating~episodes.

Measurements of the permittivity of ice and dust pebbles would be helpful to investigate the possibility to differentiate between volatiles and super-volatiles within comets and to compare observations with thermophysical modelling of the radiogenic heating. However, this is a complex endeavour because deep-penetrating radar requires the usage of large wavelengths. DC geo-electric methods with high penetration depths require spacecraft missions with at least two~landers. 

Further measurements of temperatures at the surface as well as in the upper layers of cometary nuclei are needed to constrain the makeup of the cometary subsurface through comparison with thermophysical models. However, as~there are many factors affecting temperature, such as surface roughness, self-heating and shadowing, the~models need to be applied with care. In~general, the~cooling phase in the cometary night seems to be most promising because the above influences are reduced, and~models have shown that much information can be~extracted. 

{Future missions to and astronomical observations of comets may be able to test the following most crucial predictions of the pebble-cloud-collapse model of planetesimals and subsequent evolutionary processes into~comets:
\begin{itemize}
    \item[i] Strong positive correlation between the surface temperature at sunrise and the insolation at local noon for a subsurface made of pebbles, in~contrast to no such dependency for a makeup without large void spaces~\cite{Bischoff.2021}, measurable by thermal IR mapping. Such measurements would deliver, from~remote observations only, invaluable information about the presence and size of pebbles in a shallow subsurface layer.
    \item[ii] Proof of the absence or presence of internal volatile differentiation of comets, as~predicted by Malamud et al. 2022 \cite{Malamud.2022} for comets consisting of pebbles, measurable by long-wavelength radar. Such measurements would deliver information about the formation time and/or abundance of radiogenic nuclei and about the thermal conductivity in the deep interior of the comet nucleus, and~thus would confirm or disprove the gravitational-collapse theory.
    \item[iii] Search for traces of the collisional evolution of cometary nuclei, measurable by radar through the amplitude and length scale of internal inhomogeneities. Due to the obvious distinction of the internal makeup of bodies of type A1, B1, A2, B2 and A/B3 (see Figure~\ref{fig:overview}), deep-penetrating radar measurements covering the entire body would provide information about the internal distribution of water ice, the~refractory component and the void spaces inside the body.
    \item[iv] Determination of the physical properties and orbital parameters of extinct comets. Due to memory effect (see \citep{Blum.2014}) or differentiation (see \citep{Malamud.2022}), the~depth to which dust activity is possible strongly depends on the original size of the comet-precursor body, so larger comets should be going extinct on a different timescale than smaller ones (see \citep{Gundlach.2016}).
    \item[v] Determine whether positive relief features~\cite{Davidsson.2016} are local remnants of impacts, measurable by high-frequency radar through local permittivity enhancement. If~moderate-velocity impacts on the surface of a planetesimal lead to local compaction of the material and thus to loss of the pebble structure, further dust activity is impossible due to considerable enhancement of the tensile strength. Such measurements would deliver information about the collision history of the planetesimals and the abundance of small-scale (metre- to decimetre-sized) objects in the region in which the planetesimals resided throughout most of their lifetime before becoming comets.
\end{itemize}
}

\color{black}

It should be noted that the picture of formation and evolution of planetesimals drawn in this review cannot be complete because it ignores evolutionary processes that are usually summarised under space weathering. We again emphasise that we only provide one plausible, but~not the only possible scenario of the formation of planetesimals and evolution into comets. {Advantages of the pebble-collapse model of comet formation are that it is entirely based on a large body of empirical evidence from the laboratory as well as on established numerical simulations of hydrodynamic processes. The~pebble-collapse model explains a number of observations of comets, particularly of comet 67P, such as the low bulk density, ultra-low mechanical strength, low thermal conductivity, presence of ultra-low-density dust particles and internal homogeneity. However, there are observations of comet 67P that the model cannot explain. Among~these are the observed layering, the~presence and variety of macroscopic ($\gtrsim$$1 \, \mathrm{m}$) geologic features and the measured increase of permittivity in a shallow ($\lesssim$$25 \, \mathrm{m}$) subsurface region, for~which alternative formation models may have explanations (unless they are evolutionary processes). These models are discussed in Weissman et al. 2020 \cite{Weissman.2020}.}

\section*{Acknowledgments}
The picture drawn in this review is the result of decades of research by numerous scientists whom we would like to thank. We also acknowledge that without the continuous support of funding agencies, foremost the Deutsche Forschungsgemeinschaft DFG and the Deutsches Zentrum für Luft- und Raumfahrt DLR, this work would not have been possible. This research was funded by DFG grant number BL 298/27-1.

\bibliographystyle{unsrt}  
\bibliography{references.bib}  

\end{document}